\documentclass{article}
\usepackage{amsmath,amsfonts,amssymb}
\usepackage{graphicx,epsfig} 
\numberwithin{equation}{section}
\newcommand{\ket}[1]{\left\vert #1  \right\rangle}
\newcommand{\average}[1]{\langle #1 \rangle}
\newcommand{\bra}[1]{\left\langle #1  \right\vert}
\newcommand{\kernel}[3]{\left\langle #1\left\vert #2\right\vert#3 \right\rangle}
\newcommand{\half}{\frac{1}{2}}

\newcommand{\E}{{\cal E}}

\newcommand{\Tr}{\mathrm{Tr\,}}
\newcommand{\tr}{\mathrm{tr\,}}
\newcommand{\sgn}{\mathrm{sgn\,}}
\renewcommand{\Re}{{\rm Re}}
\renewcommand{\Im}{{\rm Im}}

\def\mathbb#1{{\bf #1}}
\newtheorem{thm}{Theorem}[section]

\newtheorem{rem}[thm]{Remark}

\title{Transport and Dissipation in Quantum Pumps}
\begin{document}
\author{ J.~E.~Avron \footnote{Department of Physics, Technion, 32000 Haifa,
Israel}, A. Elgart \footnote{Courant Institute of Mathematical Sciences,
New York, NY 10012, USA}, G.M. Graf \footnote{Theoretische
Physik, ETH-H\"onggerberg, 8093 Z\"urich, Switzerland} and L.
Sadun \footnote{Department of Mathematics, University of Texas,
Austin Texas 78712, USA}}
\maketitle
\begin{abstract}
This paper is about adiabatic transport in quantum pumps. The
notion of ``energy shift'', a self-adjoint operator dual to the
Wigner time delay, plays a role in our approach: It determines the
current, the dissipation, the noise and the entropy currents in
quantum pumps.  We discuss the geometric and topological content
of adiabatic transport and show that the mechanism of Thouless and
Niu for quantized transport via Chern numbers cannot be realized
in quantum pumps where Chern numbers necessarily vanish.
\end{abstract}
\tableofcontents

\section{Introduction}
\begin{figure}[h]
\hskip 1.2 in
\includegraphics[width=4cm]{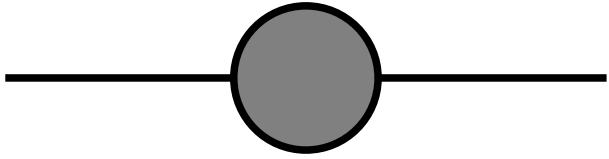}
\caption{A model of a quantum scatterer with two leads.}
\label{fig:pump}
\end{figure}

An adiabatic quantum pump \cite{ref:marcus} is a time-dependent
scatterer connected to several leads. Fig.~\ref{fig:pump} is an
example with two leads. Each lead may have several channels. The
total number of channels, in all leads, will be denoted by $n$.
Each channel is represented by a semi-infinite, one dimensional,
(single mode) ideal wire. We assume that the particles propagating
in the channels are non-interacting\footnote{Quantum pumps have
also been discussed in the context of Luttinger liquids, see e.g.
\cite{ref:aleiner, ref:sc}.} and all have dispersion $\epsilon(k)$.
For the sake of concreteness we shall take quadratic dispersion,
$\epsilon(k)=k^2/2$, but most of our results carry over to more
general dispersions. We also assume that the incoming particles
are described by a density matrix $\rho$ common to all channels:
$\rho(E)=(1+e^{\beta(E-\mu)})^{-1}$, with chemical potential $\mu$
and temperature $T$. The scatterer is adiabatic when its
characteristic frequency $\omega \ll 1/\tau$, with $\tau$ a the
typical dwell time in the scatterer. We take units so that
$k_B=\hbar=m=e=+1$.

An incoming particle sees a quasi-static scatterer. The
scattering can therefore be computed, to leading order, by
time-independent quantum mechanics, using the scattering
Hamiltonian in effect at the time that the particle reaches the
scatterer. In other words, we pretend that the Hamiltonian always
was, and always will be, the Hamiltonian seen by the particle at
the time of passage. This gives the ``frozen $S$-matrix''. Since
time-independent systems conserve energy, so does the frozen
$S$-matrix. We denote by $S(E,t)$ the on-shell frozen $S$-matrix.

The outgoing states have, to order $\omega^0$, the same occupation
density as the incoming states (since $S(E,t)$ is a unitary
$n\times n$ matrix and the incoming densities are the same for all
channels). This implies no net transport. However, at order
$\omega$, there is an
interesting interference effect: An incoming particle of well-defined
energy does not
have a well-defined time of passage. This spread in time is a
consequence of the uncertainty principle, and is not related to
the dwell time of the particle in the scatterer. Thus, even if
$\omega \ll 1/\tau$, the (frozen) $S$-matrix seen by the tail of
the wave packet will differ slightly from that seen by the head of
the wave packet. This differential scattering causes the outgoing
occupation densities to differ from the incoming densities to
order $\omega^1$, leading to a nonzero transport.


The density matrix for the outgoing states, $\rho_{out}$, is
determined by $\rho(H_0)$, the density matrix of the incoming
states, and the S-matrix. Let $S_d$ be the exact (dynamical)
S-matrix. As we shall explain in section~\ref{sec:ad-scattering},
$\rho_{out}$ is given by
 \begin{equation}
 \rho_{out}=\rho(H_0-{\E } _d),\quad
{\E } _d= i\, \dot S_d S_d^*.
\label{eq:rho-out}
\end{equation}
A dot denotes derivative with respect to time\footnote{The time
dependence of $S_d$ is discussed in section
\ref{sec:ad-scattering}.}. $\E_d$ is the operator of energy shift
introduced in \cite{ref:MartinSassoli}. It combines information
on the state of the scatterer with its rate of change $\dot S_d$.
For time independent scattering, ${\E } _d=0$ and
Eq.~(\ref{eq:rho-out}) is an expression of conservation of energy.
The formula for $\rho_{out}$, Eq.~(\ref{eq:rho-out}), holds
independently of whether the scattering is adiabatic or not.

We call the frozen analog of the operator $\E_d$ the {\em matrix}
of energy shift. It is the $n\!\times n$ matrix
\begin{equation}\label{eq:e-shift-matrix}
\E (E,t)= i\, \dot S(E,t) S^*(E,t),
\end{equation}
and is a natural dual to the more familiar Wigner time delay \cite{ref:Wigner}
\begin{equation}\label{eq:t-matrix}
{\cal T}(E,t)=-i\, S^\prime (E,t) S^*(E,t).
\end{equation}
where prime denotes derivative with respect to $E$. As we shall
see in section \ref{sec:curvature} their commutator
\begin{equation}\label{eq:omega-def}
\Omega=i[{\cal T},\E]
\end{equation}
has a geometric interpretation of a curvature in the time-energy
plane, analogous to the adiabatic curvature.

Adiabatic transport can be expressed in terms of the matrix of
energy shift. For example, the BPT \cite{ref:bpt} formula for the
expectation value of the current in the $j$-th channel,
$\average{\dot Q}_j$ takes the form:
\begin{equation}
\average{\dot Q}_j(t) = -{1\over 2 \pi}\, \int_0^\infty dE
\,\rho'(E) \,{\E}_{jj}(E,t). \label{eq:BPT}
\end{equation}
At $T=0$, $-\rho'$ is a delta function at the Fermi energy and the
charge transport is determined by the energy shift at the Fermi
energy alone.

There are two noteworthy aspects of this formula. The first one is
that $\average{\dot Q}$, which is of order $\omega$, can be
accurately computed from frozen scattering data which is only an
$\omega^0$ approximation. The second one is that the formula holds
all the way to $T=0$, where the adiabatic energy scale $\omega$ is
large compared to the energy scale $T$.

The energy shift also determines certain transport properties that
are of order $\omega^2$. An example is dissipation at low
temperatures. Let $\average{\dot E}_j(t)$ be the expectation value
of energy current in the $j$-th channel. Part of the energy is
forever lost as the electrons are dumped into the reservoir. The
part that can be recovered from the reservoir, by reclaiming the
transported charge, is $\mu\average{\dot Q}_j(t)$. We therefore
define the dissipation in a quantum channel as the difference of
the two. As we shall show in section \ref{sec:dissipation} the
dissipation at $T=0$ is\footnote{For related results on
dissipation see e.g. \cite{ref:MokbaletButt1}. For relations
between dissipation and the S-matrix see \cite{ref:akkermans}}:
\begin{equation} \average{\dot E}_j(t)-\mu\average{\dot
Q}_j(t)
=\frac{1}{4\pi}\big({\E } ^2\big)_{jj}(\mu,t)\ge 0.
\label{eq:dissipation}
\end{equation}
Both the dissipation and the current admit transport formulas that
are local in time at $T=0$. Namely, the response at time $t$ is
determined by the energy shift at {\em the same time}. This is
remarkable for at $T=0$ quantum correlations decay slowly in time
and one may worry that transport at time $t$ will retain memory
about the scatterer at early times. This brings us to transport
equations which admit a local description only at finite
temperatures.

The entropy and noise currents are defined as the difference
between the outgoing (into the reservoirs) and incoming entropy
(or noise) currents. Namely,
\begin{equation}
\dot s_j(t,\mu,T)=\dot s(\rho_{out,j})-\dot s(\rho_j),
\quad \dot s(\rho)=\frac{1}{2\pi}\int dE\,(h\circ\rho)(E,t)
\label{eq:entropy-current}
\end{equation}
where \cite{ref:Imry, ref:buttiker}
\begin{equation}\label{eq:h-def}
h(x)=
  \begin{cases}
    -x\log x-(1-x)\log(1-x) & \text{entropy}, \\
    x(1-x) & \text{noise}.
  \end{cases}
\end{equation}
In the adiabatic limit, $\omega \to 0$, and for $\omega\ll T\ll
\sqrt{\omega/\tau}$ we find (see section \ref{sec:entropy-noise})
\begin{equation}
\dot s_j(t,\mu,T)= \frac{\beta}{2\pi k}\,\Delta\E^2_j(\mu,t)\ge
0,\quad k=\begin{cases}
  2 & \text{entropy}; \\
  6 & \text{noise},
  \end{cases}
\label{eq:entropyflow}
\end{equation}
where
\begin{equation}\label{eq:delta-E-2}
 \Delta\E^2_j= \big({\E
} ^2\big)_{jj} - \big({\E }_{jj} \big)^2.
\end{equation}
When $T\lesssim\omega$ the entropy and noise currents at time $t$
are mindful of the scattering data for earlier times and there
are no transport equations that are local in time. What sets them
apart from the current and the dissipation is the {\em non-linear}
dependence on the density. The non-linearity makes the transport
sensitive to the slow decay of correlations.

Our result about the noise overlap with results that follow from
the ``full counting statistics'' of Levitov et.\ al.\
\cite{ref:LLL}. When there is overlap, the results agree. However,
the results are mostly complementary, a reflection of the fact
that both the questions and the methods are different. ``Counting
statistics'' determine transport in a {\em pump cycle} in terms of
the {\em entire history} of the pump. We give information that is
local in time. In this sense, we give stronger results. On the
other hand, the counting statistics determine {\em all} moments
all the way down to zero temperature, while our results go down to
$T=0$ for the current and dissipation but not for the entropy
and noise. A detailed comparison of our results with results that
follow from the Lesovik-Levitov formalism \cite{ref:LLL} is made in appendix
\ref{sec:levitov}. For other results on noise in pumps see e.g.,
\cite{ref:PoliankiBrouwer}.

Transport in adiabatic scattering is conveniently described using
semi-classical methods \cite{ref:semiclassics} a.k.a.
pseudo-differential (Weyl) calculus \cite{ref:robert}. As we shall
explain in section \ref{sec:ad-scattering}, $S(E,t)$ is the
principal symbol of the exact S-matrix, $S_d$. Semi-classical
methods can be used to derive
Eqs.~(\ref{eq:BPT}, \ref{eq:dissipation}, \ref{eq:entropy-current}).
For an alternate point of view using coherent states see \cite{ref:aegs:JMP}.

In section \ref{sec:geometry-topology} we discuss the geometric
and topological significance of our results. We shall see that charge
transport can be formulated in terms of the curvature, or Chern
character, of a natural line bundle. This is reminiscent of works
of Thouless and Niu \cite{ref:thouless83} which identified quantized
charge transport with Chern numbers \cite{ref:thoulessbook} and inspired
the study of quantum pumps. Nevertheless the situation is different here,
since the bundle is trivial and, besides, the integration manifold has a
boundary.
This does not preclude the possibility that charge is
quantized for reasons other than being a Chern number. In fact,
one can geometrically characterize a class of periodic pump
operations \cite{ref:AEGS:PRL} for which the transported charge in a cycle,
and not just its expectation value, is a non-random integer.
It is to be cautioned that a small change of the scattering matrix will
typically destroy this quantization.

\section{Pedestrian derivation of BPT}
\label{sec:bpt}

At $T=0$ the Fermi energy is a step function and
$\rho^\prime(E)=-\delta(\mu-E)$, and BPT, Eq.~(\ref{eq:BPT}),
takes the form
\begin{equation}
\average{\dot Q}_j (\mu,t)= {1\over 2 \pi}\, \E_{jj}(\mu,t).
\label{eq:BPT:T=0}
\end{equation}
In this section we shall describe an argument \cite{ref:AEGS:PRB}
that explains this equation in the two channel case. The two
channel case is special in that the changes in the scattering
matrix break into elementary processes so that for each one BPT
follows either from simple physical arguments or from known
facts\footnote{We assume that the transported charge depends only
on $S(\mu)$ and, linearly, on $dS(\mu)$, regardless of the
physical realization of the scatterer.}.

\subsection{The two channels case}

In the two channel case, Fig.\ \ref{fig:pump}, the frozen, on
shell, S-matrix takes the form
\begin{equation}S(\mu)=\begin{pmatrix}
  \mathrm{r} & \mathrm{t}'\\
\mathrm{t}& \mathrm{r}'
\end{pmatrix}(\mu),\end{equation} with $\mathrm{r}$ and $\mathrm{t}$
(respectively $\mathrm{r}'$ and $\mathrm{t}'$) the reflection and
transmission coefficients from the left (right).
Eq.~(\ref{eq:BPT:T=0}) reads \begin{equation} \label{bptrt} 2\pi\,
\average{dQ}_- =   i (\mathrm{\bar r dr + \bar t' dt'}),\quad
2\pi\, \average{dQ}_+ =   i (\mathrm{\bar r' dr' + \bar t dt})
\end{equation}
$\average{dQ}_-$ is made from data ($\mathrm{r},\,\mathrm{t'}$)
describing the scattering to the left, and similarly
$\average{dQ}_+$ from those to the right\footnote{This is why
$S^*$ is on the right in the energy shift,
Eq.~(\ref{eq:e-shift-matrix}).}.

To identify the physical interpretation of the differentials
we introduce new coordinates $(\theta, \alpha, \phi, \gamma)$.
The most general unitary $2 \times 2$ matrix can be expressed in
the form:
\begin{equation}
S = e^{i\gamma}\begin{pmatrix}
e^{i \alpha}\,\cos\theta  & i e^{-i\phi}\,\sin\theta  \\
i e^{i\phi}\,\sin\theta  &e^{-i \alpha}\, \cos\theta
\end{pmatrix} \label{S1} \end{equation} where $0\le \alpha,\phi<2\pi,\
0\le\gamma<\pi$ and $0\le\theta\le\pi/2$. In terms of these
parameters, the BPT formula reads \begin{equation} 2 \pi\,
\average{dQ}_\pm= \pm \left(\cos^2\theta\right)\,
d\alpha\mp\left(\sin^2\theta\right)\,d\phi - d\gamma. \label{dq2}
\end{equation}
As we shall now explain, the variations $d\alpha$, $d\phi$ and
$d\gamma$ can be identified with simple physical processes.
\subsection{The snowplow}
\label{subse:snowplow}
\begin{figure}[h]
\hskip 1.2 in
\includegraphics[width=9cm]{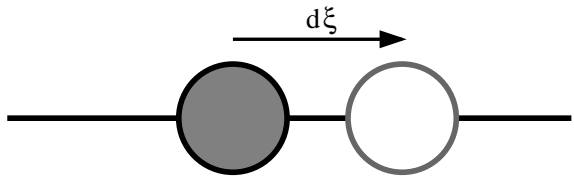}
\caption{Moving the scatterer changes $\alpha \to \alpha + 2
k_F\, d \xi$} \label{plowfigure}
\end{figure}

Let $k_F$ be the Fermi momentum associated with $\mu$. Translating
the scatterer a distance $d\xi$ multiplies $\mathrm{r,\ (r')}$ by
$e^{2ik_Fd \xi},\ (e^{-{2ik_F d \xi}}) $, and leaves $\mathrm{t}$
and $\mathrm{t'}$ unchanged. It follows that $d\alpha={2k_F d
\xi}$ corresponds to shifting the scatterer.

As the scatterer moves, it attempts to push the $k_F d\xi/\pi =
d\alpha/2\pi$ electrons that occupy the region of size $d\xi$ out
of the way, much as a snowplow attempts to clear a path on a
winter day. Of these, a fraction $\mathrm{|t|}^2=\sin^2\theta$
will pass through the scatterer (or rather, the scatterer will
pass through them), while the remaining fraction
$\mathrm{|r|}^2=\cos^2\theta$ will be propelled forward, resulting
in net charge transport of
\begin{equation} 2\pi\, \average{dQ}_\pm =\pm
\left({\cos^2\theta }\right)\, d\alpha, \label{eq:dalpha}
\end{equation}
in accordance with Eq.~(\ref{dq2}).
\begin{rem}
It is instructive to examine the special case of a uniformly
moving scatterer where we can use Galilei transformations to
compute the charge transport exactly. By taking the limit of
slowly moving scatterer we get a check on the result above.

Since the mass of the electron is one, the Galilean shift from
the lab frame to the frame of the scatterer shifts each momentum by $-\dot\xi$.
In the lab frame, the incoming states are filled
up to the Fermi momentum $k_F$ while in the moving frame the
incoming states of the $\pm$ channels are filled up to
momenta $k_F\pm\dot\xi$. In the moving frame, the outgoing states
on the $\pm$ channels are filled up to $k_F\mp\dot\xi$, and
partially filled with density $|\mathrm{t}'(k)|^2$ for momenta in
the interval $(k_F-\dot\xi, k_F+\dot\xi)$. Transforming back to
the lab frame we find for $\delta\rho$ of the $-$ (=left) channel
\begin{equation}\delta\rho_{-}(k^2)=
  \begin{cases}
    0 & \text{if $k<k_F - 2\dot\xi$} \\
  - |\mathrm{r}'(k+\dot\xi)|^2  & \text{if $k_F-2\dot\xi < k < k_F$} \\
   0 & \text{if $k>k_F$}.
  \end{cases}
\end{equation}
To order $\dot\xi$,
\begin{equation}\label{eq:delta-rho}
\delta\rho_{-}(E)=-2k_F\dot\xi\,|\mathrm{r}'(k_F)|^2\delta(E-\mu).
\end{equation}
Since the current is
\begin{equation}\label{eq:current-heuri} \average{\dot Q}_j(t)=
\frac 1 {2\pi}\int_0^\infty\,dE\, \,\delta\rho_{j}(E,t),
\end{equation}
Eq.~(\ref{eq:dalpha}) is reproduced.
\end{rem}\begin{rem}
The net outflow of charge, $\average{dQ}_- + \average{dQ}_+$,
vanishes to order $\dot\xi$ but not to order $\dot\xi^2$. This is
because the moving snowplow leaves a region of reduced density in
its wake.
\end{rem}

\subsection{The battery}
\label{subse:bat}

\begin{figure}[h]
\hskip 1.4 in
\includegraphics[width=7cm]{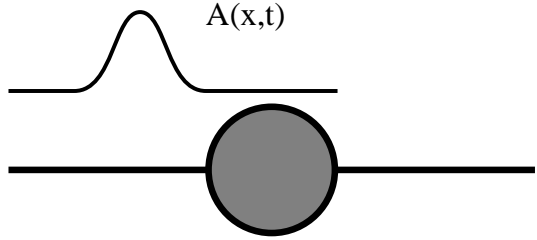}
\caption{Applying a vector potential changes $\phi \to \phi +
\int A$}\label{voltfigure}
\end{figure}

To vary $\phi$ we add a vector
potential $A$. This induces a phase
shift $d\phi = \int A$ across the scatterer, and multiplies
$\mathrm{t,\ (t')}$ by $e^{i d \phi},\
(e^{-i d \phi}) $, while leaving $\mathrm{r}$ and $\mathrm{r'}$ unchanged.
This phase shift depends only on $\int A$, and
is independent of the placement or form of the
vector potential.
The variation in the vector potential induces an EMF of
strength $ -\int \dot A = -\dot\phi$. To first order,
the current is simply the
voltage times the Landauer conductance $\mathrm{|t|}^2/2\pi$
\cite{ref:Imry}. That is,
\begin{equation}
2\pi\,\average{dQ}_\pm =\mp \left(\sin^2\theta\right)\, d\phi. \label{dphi}
\end{equation}
in agreement with Eq.~(\ref{dq2}).
\begin{rem}
Consider the special case of a time independent voltage drop. In a
gauge where the battery is represented by a scalar potential, the
pump is represented by a time independent scattering problem where
the potential has slightly different asymptotes at $\pm\infty$. If
the battery is placed to the left of the scatterer, the states of
particles incident from that side are occupied up to energy
$\mu-\dot\phi$, while those incident from the right are occupied
up to energy $\mu$. Suppose $\dot\phi$ is negative. Then
$\delta\rho_+$ is
\begin{equation} \delta\rho_{+}(E) =
  \begin{cases}
    0 & \text{if $E < \mu$}, \\
    |\mathrm{t}(E)|^2  & \text{if $\mu < E < \mu-\dot\phi $ },\\
0 & \text{if $E > \mu-\dot\phi$}
  \end{cases}
\label{outbat1} \end{equation} If, however, the battery is placed
to the right of the scatterer then $\delta\rho_+$ is
\begin{equation} \delta\rho_{+}(E) =
  \begin{cases}
    0 & \text{if $E < \mu$}, \\
    |\mathrm{t}(E+\dot\phi)|^2  & \text{if $\mu < E < \mu-\dot\phi $ },\\
0 & \text{if $E > \mu-\dot\phi$}
\label{outbat2}
  \end{cases}\end{equation}
In either case, to leading order in $\dot\phi$,
\begin{equation}\label{eq:delta-rho-E}
\delta\rho_+(E)=-\dot\phi\, |\mathrm{t}(\mu)|^2\delta(E-\mu).
\end{equation}
Plugging in Eq.~(\ref{eq:current-heuri}), we recover
Eq.~(\ref{dq2}).
\end{rem}
\begin{rem}
To leading order in $\dot\phi$, $\delta\rho_+$ is independent of
whether the battery is to the left of the pump or to the right of
it. To order $\dot\phi^2$ this is no longer true as one sees from
Eq.~(\ref{outbat1}, \ref{outbat2}). The frozen S-matrix is,
however, insensitive to the location of the battery. It follows
that it is impossible to have a formula for $\delta\rho$, accurate
to order $O(\omega^2)$, that involves only the frozen $S$-matrix
and its derivatives (see \cite{ref:MoskaletButt2} for some model
calculations in the non-adiabatic regime).
\end{rem}

\subsection{The sink}
\begin{figure}[h]
\hskip 1.4in
   \includegraphics[width=7cm]{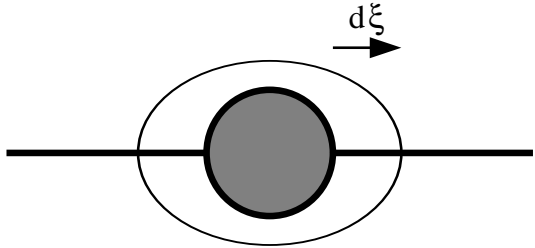} 
\caption{A scatterer that has gobbled up $d\xi$ of each wire.}\label{sink}
\end{figure}

The scattering matrix depends on a choice of fiducial points: The
choice of an origin for the two channels. Moving the two fiducial
points out a distance $d\xi$ may be interpreted as forfeiting part
of the channels in favor of the scatterer. This new, bigger,
scatterer is shown schematically in Fig. \ref{sink}. This
transforms the scattering matrix according to
\begin{equation}
S(k_F)\to e^{id\gamma}\,S(k_F),\quad d\gamma= 2k_F d\xi.
\end{equation}
This operation removes $k_Fd\xi/\pi$ electrons from each channel
and so we get
\begin{equation}
2\pi \,\average{dQ}_\pm = -2k_Fd\xi=-d\gamma,
\label{eq:dgamma}
\end{equation}
in accordance with Eq.~(\ref{dq2}). Changing $\gamma$ is therefore
equivalent to having the pump
swallow particles from the reservoirs.

For arbitrary variations $dS$ the above result still holds for the
sum $dQ_-+dQ_+$. This follows directly from a fact in scattering
theory known as Birman-Krein formula \cite{ref:krein} and in
physics as Friedel sum rule \cite{ref:friedel} which says that the
excess number of states below energy $\mu$ associated with the
scatterer is $(2\pi i)^{-1}\log\det S(\mu )$, whence
\begin{equation}
-2\pi\,\big(\average{dQ}_- + \average{dQ}_+\big)(\mu) = -id\log\det S(\mu )=2d\gamma.
\label{eq:dgamma-krein}
\end{equation}
\subsection{The ineffective variable}

We have already seen that changes in $\alpha,\phi,\gamma$ yield
transported charges $dQ_\pm$ which are correctly reproduced by
Eq.~(\ref{dq2}). Moreover, for any change of $s$, the sum
$\average{dQ}_-+ \average{dQ}_+$ is, too. To complete the
derivation of Eq.~(\ref{dq2}) we must consider variations in
$\theta$, with $\alpha,\phi$ and $\gamma$ fixed, and show that
$\average{dQ}_- - \average{dQ}_+=0$.

\begin{figure}[h]
\hskip .8 in
\includegraphics[width=8cm]{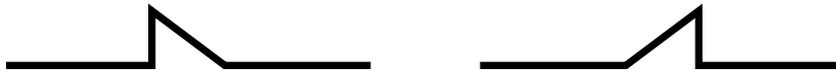}
\caption{An asymmetric scatterer and its image under
reflection}\label{mirror:fig}
\end{figure}

Suppose a scatterer has $\alpha,\phi$ and $\gamma$ fixed, but
$\theta$ changes with time. By adding a (fixed!) vector
potential and translating the system a (fixed!)
distance\footnote{This can be achieved, in general, only at a
fixed energy, and we pick the energy to be the Fermi energy
$\mu$.}, we can assume that $\alpha=\phi=0$. Now imagine a second
scatterer that is the mirror image of the first (i.e., with right
and left reversed) as in Fig. \ref{mirror:fig}. Since $\theta$ and
$\gamma$ are invariant under right-left reflection, and since
$\alpha$ and $\phi$ are odd under right-left reflection, the
second scatterer has the same frozen $S$ matrix as the first, and
this equality persists for all time.

>From the frozen scattering data we therefore conclude that the
currents for the second system are the same as for the first.
However, by reflection symmetry, $\average{dQ}_--\average{dQ}_+$
for the first system equals $\average{dQ}_+-\average{dQ}_-$ for
the second. We conclude that $\average{dQ}_+-\average{dQ}_-=0$ for
variations of $\theta$.
\section{Alternative perspectives on BPT}
The pedestrian argument does not extend beyond the two channel
case. This is because with more than two channels, a general
variation $dS$ cannot be described in terms of known physical
processes. One can, nevertheless, derive BPT from general and
simple physical considerations, without recourse to formal
perturbation expansions in scattering theory.

\subsection{An axiomatic derivation}
The BPT formula for the current follows from the following natural axioms:

\begin{enumerate}
\begin{item} Existence and Bilinearity:
\begin{equation}
dQ_m = \sum_{ijkl} a_{ijkl}^m dS_{ij} \bar S_{kl} + \hbox{ complex
conjugate}, \label{bil}
\end{equation}
with universal (complex) coefficients $a_{ijkl}^m$.
\end{item}
\begin{item} Covariance: The formula is covariant under permutation of
the channels. In particular, it is invariant under permutations
of the channels other than the $m$-th.
\end{item}

\begin{item} Gauge invariance: The formula is unchanged by time-independent
gauge transformations, and also under time-independent changes in
the fiducial points.
\end{item}

\begin{item} Cluster: If the system consists of two subsystems, disconnected
from one another, then the currents in each subsystem depend only
on the part of the $S$ matrix that governs that subsystem.
\end{item}

\begin{item} Landauer: If a voltage, applied to a single lead $m'\neq m$,
is modeled by a time-dependent gauge transformation, then $dQ_m$
is given by the Landauer formula where the transmission
probability is given by the scattering probability $m'\to m$.
\end{item}

\begin{item} Birman-Krein: $$\sum_j dQ_j =
{i \over 2\pi} d\log(\det S)={i \over 2\pi}\Tr(dS S^\dagger).$$
\end{item}
\end{enumerate}

The physical motivations for most of the axioms are clear. For
example, bi-linearity comes from the fact that the current is an
interference effect between the original outgoing wavefunction
(described by $S$) and an additional piece (described by $dS$).
The one axiom that seems the most arbitrary is existence. Namely,
the assumption that charge transport at $T=0$ is determined by the
scattering matrix at the same time and at the Fermi energy alone.

By covariance, it suffices to study the current on the first
channel, $dQ_1$ (and drop the superindex in $a$). Henceforth,
Latin indices will run from 1 to $n$, while Greek indices will run
from 2 to $n$.

The most general bilinear (\ref{bil}) vanishing identically on
unitaries $S$ and their variations $dS$ is of the form
$a_{ijkl}=e_{jl}\delta_{ik}+d_{ik}\delta_{jl}$ with hermitian
matrices $(e_{jl})$ and $(d_{ik})$. Indeed, by $SS^*=S^*S=1$ the
matrices $(dS)S^*$ and $(dS)^*S$ are anti--hermitean, which
implies vanishing of the stated bilinear form. We can thus fix
$a_{ijkl}$ in (\ref{bil}) by imposing the uniqueness constraints
that $\sum_\alpha a_{\alpha j\alpha l}$ and $\sum_\alpha
a_{i\alpha k\alpha}$ are anti--hermitian.

We shall now see that by axiom 3
\begin{equation}\label{cjk}
a_{ijkl}=c_{ij}\delta_{ik}\delta_{jl}
\end{equation}
Indeed, if we move the fiducial point on the $i$-th channel by
distance $\xi_i$, the $S$-matrix transforms by
\begin{equation} S_{ij} \to S_{ij}
e^{i k_F (\xi_i + \xi_j)}.  \end{equation} and so $a_{ijkl}\to
a_{ijkl}e^{i k_F (\xi_i + \xi_j-\xi_k-\xi_l)}$. The only invariant
terms are those with $i=k$ and $j=l$ or with $i=l$ and $j=k$.
Similarly, gauge transformations send
\begin{equation} S_{ij} \to S_{ij}
e^{i ({\phi}_i - {\phi}_j)}.  \end{equation} Now the invariant
$a_{ijkl}$ terms are those with $i=k$ and $j=l$ or with $i=j$ and
$k=l$. Putting the two selection rules together gives
Eq.~(\ref{cjk}). At this point Eq.~(\ref{bil}) reduces to
\begin{equation}
dQ_1 = \sum_{ij} c_{ij} dS_{ij} \bar S_{ij} + \hbox{ complex
conjugate}, \label{bil1}
\end{equation}
with the uniqueness constraints that $c_{ij}$ are pure imaginary.

By the clustering property $c_{\alpha\beta}$ must all vanish, for
otherwise $dQ_1$ will be affected also by channels disconnected
from it. Now $c_{1\alpha}=c_2$ and $c_{\alpha1}=c_3$ independent
of $\alpha$ by permutation symmetry. We can therefore write
Eq.~(\ref{bil}) as
\begin{equation} \label{c5}
dQ_1 =  c_1 dS_{11} \bar S_{11} + c_2 \sum_{k} dS_{1k}\bar S_{1k}
+ c_3 \sum_k dS_{k 1}\bar S_{k 1} + \hbox{ complex conjugate},
\end{equation}
Summing $dQ_j$ over all channels, using that $c_1,\,c_2,\,c_3$ are
pure imaginary, and setting the result to agree with the
Birman-Krein formula gives
\begin{equation} \sum_j dQ_j =
2ic_1\Im \sum_j dS_{jj} \bar S_{jj} + 2(c_2+c_3)\Tr(dS
S^\dagger)={i \over 2\pi}\Tr(dS S^\dagger).
\end{equation}
Thus $c_1=0$ and $4\pi(c_2+c_3)=i$.

What remains is to distinguish between $c_2$ and $c_3$. Imagine
modeling a voltage $V$ on channel $\alpha$ by a time dependent
vector potential that shifts the phase of the wavefunction on the
$\alpha$-th channel by $d\phi$. Equating with Landauer gives
\begin{equation}
dQ_1 = -2i (c_2 |S_{1\alpha }|^2 - c_3 |S_{\alpha 1}|^2)\,d\phi=
\frac 1 {2\pi}\, |S_{1\alpha }|^2 d\phi.\end{equation}
Since (for $n\ge 3$) $|S_{\alpha 1}|^2$ and $|S_{1 \alpha}|^2$ are
independent, this implies that $c_2 = i/4\pi$ and $c_3=0$. We
thus obtain BPT for $n\ge 3$\footnote{To get the $n=2$ case,
consider a 2-channel scatterer as a degenerate 3-channel
scatterer, where the third channel is disconnected from the first
two. Then $S_{31}$ is identically zero, and the 3-channel BPT
formula for $dQ_1$ reduces to the 2-channel formula.}.

\subsection{Classical pumps}
\label{sec:cl-pumps} The classical phase space associated to a
given channel is the half-plane $\{x,p\,|x>0,\,p\in \mathbb{R}\}$.
We can choose coordinates so that points in phase space are
labelled by the pair $(E,t)$, where $E$ is the energy of the
(classical) particle and $t$ its time of passage at the origin. For
concreteness, let us assume a dispersion
relation $\epsilon(p)=\epsilon(-p)$ with $\epsilon(p)$ increasing from
$0$ to $\infty$ as $p$ ranges over the same interval, e.g. a quadratic
dispersion. Then
\begin{equation}\label{eq:e-t-canonical} E=\epsilon(p),\quad t
=-x/v,\quad (v=\epsilon^\prime)
\end{equation}
is a canonical transformation to the energy-time half plane
$\{E,t\,|E>0,\,t\in \mathbb{R}\}$, since $dE\wedge dt=dx\wedge dp$.
The mapping is singular when $v=0$.
States with $t>0$ are
incoming (at time $0$), while those with $t<0$ are outgoing (at time $0$).
(All states are, of course, incoming in the distant past and outgoing in the
distant future.)

The phase space of $n$ disconnected channels is
$\Gamma=\cup_{i=1}^n\Gamma_i=
\{(E,t,i)\mid E>0,\,t\in\mathbb{R},\,i=1,\ldots n\}$. When analyzing pumps,
i.e., channels communicating through some pump proper, $\Gamma$ still serves
as phase space of the {\it scattering states}. More precisely, $(E,t,i)$ shall
be the label of the scattering state whose past asymptote is the free
trajectory with these initial data. Similarly, we may indicate a scattering
state by its future oriented data $(E',t',j)$. In this way we avoid
introducing the full phase space for the connected pump. However, some of
these trajectories may admit only one of the two labels, as they are free
for, say, $t\to -\infty$ but trapped as $t\to+\infty$. With this exception
made, the relation defines a bijective map, the (dynamical)
{\it scattering map\/}:
\begin{equation}\label{eq:scl}
S: \Gamma^-\to\Gamma^+,\quad (E,t,i)\mapsto(E',t',j),
\end{equation}
where $\Gamma\setminus\Gamma^-$ are the incoming labeled trajectories which
are trapped in the future, and correspondingly for $\Gamma\setminus\Gamma^+$.
If (\ref{eq:scl}) is viewed as a function of $(E,t)$ with $i$ fixed, the
channel $j$ is piecewise constant and the map to $(E',t')$ symplectic.
We shall illustrate $S$ by an example in Sect.~\ref{sec:phasespace}. The
inverse map may be written as
\begin{equation}\label{eq:Sinv}
S^{-1}: (E',t',j)\mapsto(E,t,i)=(E'-\E_d(E',t',j),t'-{\cal T}_d(E',t',j), i),
\end{equation}
which {\it defines} the classical energy shift $\E_d$ and the time
delay ${\cal T}_d$ as functions of the outgoing data. We remark that
for a given static pump $\E=0$ and ${\cal T}$ is independent of $t'$.
For adiabatic pumps, we have $\E_d(E',t',j)=O(\omega)$ and
${\cal T}_d(E',t',j)={\cal T}(E',t',j)+O(\omega)$,
where ${\cal T}$ is the time delay of the static scatterer in effect at the
time of passage $t'$, on which it then depends parametrically. Since
$S^{-1}$ is volume preserving by Liouville's theorem, and the
derivative w.r.t. time brings in a factor $\omega$, we have
\begin{equation}\label{liou}
{\E }^\prime+\dot{\cal T}=0,
\end{equation}
where $\E$ is the part of $\E_d$ of order $\omega^1$. This relation shows
that the static scattering data determine $\E(E',t',j)$ up to an additive
function of $t',\,j$ and, as we shall see, cannot do better. This is in sharp
contrast to the quantum case, where $\E$ is fully determined by the frozen
scattering matrix, see Eq.~(\ref{eq:e-shift-matrix}). We will further
comment on the origin of this ambiguity in the classical case in
Sects.~\ref{sec:scattering}, \ref{sec:nophase} and relate it to the lack
of phase information in classical scattering.

Similar to the quantum case, Eq.~(\ref{eq:BPT}), is however the expression
of the current in terms of the energy shift:
\begin{equation}\label{BPTclass}
\dot Q_j(t) = -\int_0^\infty dE
\,g'(E) \,{\E}(E,t,j),
\end{equation}
where $g(E)$ is the phase space particle density in the incoming flow. We
remark that in a semiclassical context $g$ is related to the occupation
density $\rho$ by $g(E)=\rho(E)/2\pi$.

In fact, the net outgoing charge transmitted in the time
interval $[0,T]$ through channel $j$ is
\begin{equation}\label{dQcl}
Q_j=\int_{\Gamma_j}dE'dt'\chi_{[0,T]}(t')g(E)\,-\,
\int_{\Gamma_j}dEdt\chi_{[0,T]}(t)g(E),
\end{equation}
where $E$ in the first integral is given through the map
(\ref{eq:Sinv}) if $(E',t',j)\in \Gamma^+$; if
$(E',t',j)\in\Gamma\setminus\Gamma^+$, which may occur if $E'$ is close
to threshold energy $0$ and $E<0$, we assume that $g(E)=g(0)$, i.e., that
the occupation of the bound states and threshold energies are equal.

Eq.~(\ref{eq:BPT}) is obtained immediately by expanding
$g(E)=g(E')-g'(E')\E(E',t',j)+O(\omega^2)$ in the first
integral~(\ref{dQcl}). The contribution of the first term cancels against the
second integral.

Another derivation, which is more involved, is of some interest
especially in view of the semiclassical discussion of pumps in
Sect.~\ref{sec:time-energy}. The first integral (\ref{dQcl})
equals
\begin{equation}\label{class0}
\int_{\Gamma_j}dE'dt'\chi_{[0,T]}(t)g(E)\,-\,
\int_0^\infty dE'g(E){\cal T}(E',t',j)\Big|_{t=0}^{t=T}.
\end{equation}
In the adiabatic regime we may describe $\Gamma_j$, to lowest approximation,
as $\Gamma_j=\cup_{i=1}^n\Gamma_{ij}$, where $\Gamma_{ij}$ consists of states
$(E',t',j)$ originating from lead $i$ under static scattering. W.r.t. them
and to next approximation, their preimages $(E,t,i)$ appearing as arguments
in the first
integral (\ref{class0}), are displaced by the vector field
$-(\E(E',t',j), {\cal T}(E',t',j))$, which is typically discontinuous across
the boundaries of the $\Gamma_{ij}$'s, but divergence free otherwise by
(\ref{liou}). (For an
illustration, see Example~\ref{sec:phasespace}
and Fig.~\ref{fig:phasespace} there.) As a result,
that integral differs from the second integral (\ref{dQcl}) by
\begin{equation}\label{class1}
-\sum_{i=1}^n\int_{\partial \Gamma_{ij}}
(d\sigma_E\E(E',t',j)+d\sigma_t{\cal T}(E',t',j))\chi_{[0,T]}(t)g(E),
\end{equation}
where $(d\sigma_E, d\sigma_t)$ is the outward normal to
$\partial \Gamma_{ij}$. Within $\cup_{i=1}^n\partial\Gamma_{ij}$ we may
distinguish between boundary parts contained in the boundary
$\{E=0\}$ of $\Gamma_j$, and inner boundaries. The contribution of the
former is $\int_0^T dt g(0)\E(0,t,j)$ and, mostly for comparison with the
promised semiclassical derivation, we formally write
that of the latter as
$\int_{\Gamma_j}\Omega(E,t,j)\chi_{[0,T]}(t)g(E)$,
where $\Omega(E,t,j)$ is a distribution supported on the inner boundaries.
Since the map (\ref{eq:scl}) is bijective on $\Gamma$ except for bound
states, the displacements of the sets $\Gamma_{ij}$ are such that
$\sum_{j=1}^n\Omega(E,t,j)=0$. In summary, we obtain
\begin{equation}\label{class2}
Q_j=\int_0^T dtg(0)\E(0,t,j)
-\int_0^\infty dE\,g(E)\dot{\cal T}(E,t,j)\Big|_{t=0}^{t=T}
+ \int_0^\infty dE\int_0^T dtg(E)\,\Omega(E,t,j)
\end{equation}
and, by differentiating w.r.t. $T$,
\begin{equation}\label{class3}
{\dot Q}_j= g(0)\E(0,t,j)-
\int_0^\infty dEg(E)\dot{\cal T}(E,t,j)
+\int_0^\infty dEg(E)\,\Omega(E,t,j).
\end{equation}
The first term on the r.h.s. describes the release
and trapping from bound states. The middle term describes the
depletion of the outgoing
flow as a result of a time delay increasing over time, since
effectively no charge is exiting during a time $d{\cal T}$.
The last term describes electrons that are reshuffled
between leads, but with no withholdings
since $\sum_{j=1}^n \Omega(E,t,j)=0$.

>From (\ref{class3}), Eq.~(\ref{BPTclass}) can be recovered: Let
$[E_k(t),E_{k+1}(t)],\,(k=0,1, \ldots)$ be the intervals of the partition of
$[0,\infty)$ on which $\E,\,{\cal T}$ are continuous, and
$\Delta_k\E=\E(E_k+,t,j)-\E(E_k-,t,j),\,
\Delta_k{\cal T}={\cal T}(E_k+,t,j)-{\cal T}(E_k-,t,j),\,(k=1,2, \ldots)$
the values of their
discontinuities at the endpoints. Then
\begin{gather}
\int_0^\infty dEg(E)\,\Omega(E,t,j)
=\sum_{k\ge 1}g(E_k)(\Delta_k\E-\dot{E_k}\Delta_k{\cal T}),\nonumber\\
-\int_0^\infty dEg(E)\dot{\cal T}(E,t,j)
=-\sum_{k\ge 0}\int_{E_k}^{E_{k+1}} dEg(E)\dot{\cal T}(E,t,j)
+\sum_{k\ge 1}g(E_k)\dot{E_k}\Delta_k{\cal T}.
\label{class5}
\end{gather}
Using Eq.~(\ref{liou}) and integration by parts, the first term on the
r.h.s. of (\ref{class5}) can be written as
\begin{equation*}\label{class6}
\sum_{k\ge 0}\int_{E_k}^{E_{k+1}}\! dEg(E){\E }^\prime(E,t,j)
=-g(0)\E(0,t,j)-\sum_{k\ge 1}g(E_k)\Delta_k\E
-\sum_{k\ge 0}\int_{E_k}^{E_{k+1}} \! dEg'(E){\E}(E,t,j).
\end{equation*}
By collecting terms, we recover Eq.~(\ref{BPTclass}).

\subsection{Currents and the $\cal T $-$\cal E $ uncertainty}
\label{sec:time-energy}
We present a semiclassical derivation of Eq.~(\ref{eq:BPT}), where we
however take for granted the physical meaning of energy shift of
Eq.~(\ref{eq:e-shift-matrix}), which will be established in
Sect.~\ref{sec:scattering}. Since the time delay is a matrix,
see Eq.~(\ref{eq:t-matrix}), we should, as a preliminary, point out that
it is its diagonal
element ${\cal T}_{jj}(E,t)$ which has the meaning of the average time delay
of a particle exiting channel $j$. In fact, consider an incoming wave packet
$\int dk e^{-i(kx+\epsilon(k)t)}\varphi(k)$ in channel $i$ centered on the
trajectory $-x=\epsilon'(k)t+c$. The part of it scattered into lead $j$
is $\int dk e^{i(kx-\epsilon(k)t)}S_{ji}(\epsilon(k),t)\varphi(k)$ and is
associated with $x=\epsilon'(k)(t-(\arg{S_{ji}})')+c$ (with
$'=d/dE$ on $\arg{S_{ji}}$), which implies a time
delay of $(\arg{S_{ji}})'$. Averaging this with the probability
$|S_{ji}|^2$ for the particle to have come from channel $i$, we find for the
average delay
\begin{equation}
\sum_{i=1}^n|S_{ji}|^2(\arg{S_{ji}})'=
\Im \sum_{i=1}^n\bar S_{ji}S'_{ji}={\cal T}_{jj}.
\end{equation}

The net outgoing charge in the time
interval $[0,T]$ is at order $\omega$
\begin{equation}
\average{Q}_j=\frac{1}{2\pi}\int_0^\infty dE'\int_0^Tdt'\,\rho(E)\,-\,
\frac{1}{2\pi}\int_0^\infty dE\int_0^Tdt\, \rho(E)\;, \label{dQa}
\end{equation}
where $E$ in the first integral is given through the map
\begin{equation}
\Phi: (E',t')\mapsto(E,t)=\big(E'-\E_{jj}(E',t'),t'-{\cal
T}_{jj}(E',t')\big)\;, \label{shift1}
\end{equation}
which describes the effect of the pump on the energy and the time
of passage of an electron in terms of the outgoing data $(E',t')$.
This is similar to the classical Eq.~(\ref{eq:Sinv}) except
that $\E$ and ${\cal T}$ are now defined in terms of the quantum mechanical
frozen scattering matrix $S$, see Eqs. (\ref{eq:e-shift-matrix},
\ref{eq:t-matrix}).
Energies $E'$ close to the threshold $E'=0$ may not be in the domain of
the map $\Phi$. Similarly, energies $E$ may there fail to
be in its range. These situations correspond to electrons released
from, resp.~trapped into, a bound state of the pump.

The BPT formula, Eq.~(\ref{eq:BPT}), is again obtained immediately by expanding
$\rho(E)=\rho(E')-\rho'(E')\E_{jj}(E',t')+O(\omega^2)$ in the first
integral~(\ref{dQa}). The contribution of the first term cancels against the
second integral. A further derivation, which is longer but adds another
interpretation to the result, is by using $\Phi$ as a change of variables.
The Jacobian
of (\ref{shift1}) is $1-\Omega_{jj}(E',t')$, where
$\Omega_{jj}$ is the divergence of the displacement
$(\E_{jj},{\cal T}_{jj})$. As a matrix, $\Omega$ is the
time delay-energy shift uncertainty introduced in Eq.~(\ref{eq:omega-def}):
\begin{equation}\label{eq:omega-1}
\Omega=i[{\cal T},\E]=i\,(\dot S S^{*\prime}-S'\dot{ S}^*)=
{\E }^\prime+\dot{\cal T}\;.
\end{equation}
Since $\Omega$ is formally of order $O(\omega)$, the Jacobian is
close to $1$ and the map (\ref{shift1}) is invertible. After
changing variables to $(E,t)$ the first integral (\ref{dQa})
extends over $(E,t)\in\Phi([0,\infty)\times[0,T])$, which differs
from $[0,\infty)\times[0,T]$ to leading order by displacements
$-\E_{jj}(E,t)$ along $E=0$ and $-{\cal T}_{jj}(E,t)$ along
$t=0,T$. This yields
\begin{equation*}
2\pi \average{Q}_j= -\int_0^T\!dt
\rho(E)\E_{jj}(E,t)\Big|_{0}^{\infty} \!-\int_0^\infty\!dE \rho(E){\cal
T}_{jj}(E,t)\Big|_{t=0}^{t=T}\!+ \int_0^\infty\!dE\int_0^T\!dt
\rho(E)\Omega_{jj}(E,t)\;,
\end{equation*}
and the differential version thereof is
\begin{equation}
2\pi\average{\dot Q}_j(t)=
\rho(0)\E_{jj}(0,t) -\int_0^\infty dE\,\rho(E)\dot{\cal
T}_{jj}(E,t)+ \int_0^\infty dE\,\rho(E)\,\Omega_{jj}(E,t)\;. \label{pcf}
\end{equation}
Upon inserting (\ref{eq:omega-1}) and performing an integration by
parts on $\rho\E'_{jj}$, Eq.~(\ref{eq:BPT})
is recovered. The interpretation of the three terms is the same as given after
Eq.~(\ref{class3}). For the first term, with bound states now quantized,
this is further discussed in the remark
below. As for the last term, notice that $\sum_{j=1}^n \Omega_{jj}=0$
still holds true because of Eq.~(\ref{eq:omega-def}). While generally, and in
contrast to the classical case, $\Omega$ may have full support in phase
space, it remains
true that it vanishes if scattering is deterministic: If on some open
subset of phase space
\begin{equation}
S_{ji}(E,t)=0\;\hbox{ for }\;i\neq\pi(j),
\end{equation}
where $\pi$ is some fixed permutation of the channels, then $S^*$ is
in the same manner related to $\pi^{-1}$, and $\E, {\cal T}$ to
the identity permutation, i.e., they are both diagonal matrices. Hence
$\Omega=0$ by (\ref{eq:omega-def}).\\


\begin{rem}
The first term on the r.h.s. of (\ref{pcf}) typically consists
of delta functions located at times $t$ where the pump has a
semi-bound state at a band edge, i.e., a state which can be turned
either into a bound state or a scattering state by an arbitrarily
small change of the pump configuration. We illustrate this for
$\epsilon(k)=k^2$, and first
claim: either $S(0,t)\equiv\lim_{E\downarrow 0}S(E,t)=-1$, or the
(static) pump at time $t$ admits a semi-bound state.
This is seen as follows: For each $k\in\mathbb{C}$ the
$2n$ plane waves
\begin{equation}
e_i\cos{kx}\;,\qquad e_i\frac{\sin{kx}}{k}\;, \label{plw}
\end{equation}
($\{e_i\}_{i=1}^n$ being the standard basis of $\mathbb{C}^n$),
form a basis of solutions with energy $k^2$ in the $n$
disconnected leads. Upon connecting them to the scatterer an
$n$-dimensional subspace of solutions is left, which depends
analytically on $k$. Since the dependence of (\ref{plw}) is also
analytic, solutions may be written as
\begin{equation}
\psi_k(x)=\sum_{i=1}^n\alpha_ie_i\cos{kx}+\beta_ie_i\frac{\sin{kx}}{k}\;,
\label{psik}
\end{equation}
with amplitudes
$\alpha=(\alpha_1,\ldots,\alpha_n),\,\beta=(\beta_1,\ldots,\beta_n)$
satisfying a set of linear equations
\begin{equation}
A(k)\alpha+B(k)\beta=0 \label{sub}
\end{equation}
with analytic $n\times n$ coefficient matrices $A,\,B$. As
(\ref{sub}) defines an $n$-dimensional subspace, we have
$\mathrm{rank\,}(A,B)=n$. For $k=0$ (\ref{plw}) reduce to $e_i,\,e_ix$. By
a semi-bound state we mean more precisely a bounded solution
$\psi_0(x)$, i.e., one with $\beta=0$ in (\ref{psik}). Its
existence is tantamount to $\det A(0)=0$. If $\det A(0)\neq 0$,
(\ref{sub}) can be solved for $\alpha$ at small $k$:
$\alpha=-A(k)^{-1}B(k)\beta$, with arbitrary
$\beta\in\mathbb{C}^n$. For $k>0$, the scattering matrix maps the
incoming part of (\ref{psik}) to its outgoing part,
$ik\alpha+\beta=S(k^2)(ik\alpha-\beta)$. Thus
\begin{equation}
S(k^2)=(-ikA(k)^{-1}B(k)+1)(-ikA(k)^{-1}B(k)-1)^{-1}\to
-1\;,\qquad (k\to 0)\;. \label{sab}
\end{equation}
This proves the claim, and in particular that $\E(0,t)=0$ except at times $t$
when the scatterer has a semi-bound state. To discuss its behavior
there, say this happens at $t=0$, we assume generically that
$C(k,t)=B(k,t)^{-1}A(k,t)\,(=C(k,t)^*)$ has a simple eigenvalue $\gamma(k,t)$
with a first order zero at $k=t=0$. Let $P$ be its eigenprojection at crossing
and let $\sigma=\sgn\dot\gamma(0,0)$ be the crossing direction. Then, we claim,
\begin{equation}
\lim_{E\downarrow 0}\E(E,t)\,dt=-2\pi\sigma P\delta(t)\,dt\;,
\label{bpt0}
\end{equation}
so that in the process, by (\ref{eq:BPT:T=0}), the charge
\begin{equation}
-\lim_{E\downarrow 0}\frac{1}{2\pi}\int_{-\epsilon}^{\epsilon}\tr\E(E,t)\,dt
=\sigma
\label{cap}
\end{equation}
is captured at arbitrarily small energy. Since the l.h.s. also equals
$\lim_{E\downarrow 0}\arg\det S(E,t)\vert_{-\epsilon}^{\epsilon}$, (\ref{cap})
states that the phase of $S(0,t)$ changes by $2\pi$ upon capture of a bound
state, which is a dynamical version of Levinson's theorem. The proof of
(\ref{bpt0}) rests on the approximation of (\ref{sab})
\begin{equation}
S(k^2,t)=\frac{ikP-(\gamma'(0,0)k+\dot\gamma(0,0)t)}
{ikP+(\gamma'(0,0)k+\dot\gamma(0,0)t)}\;,
\end{equation}
valid near $k=t=0$.
\end{rem}

\section{Time dependent scattering and Weyl calculus}
\subsection{The energy shift}\label{sec:scattering}

In this section we describe the notion of energy shift in the
context of time dependent scattering theory and derive an operator
identity relating the outgoing density $\rho_{out}$ to the
incoming density and the energy shift.

Energy is conserved in time independent scattering but not in time
dependent scattering. This leads to an important differences
between the S-matrix in the time independent and the time
dependent case: In the time dependent case the S-matrix acquires a
dependence on time shifts. Since this observation is not
particular to pumps, it is simpler to describe it in general
terms.

Using the conventional notation of scattering theory, let $H_0$ be
our reference, time independent, Hamiltonian (often called the
free Hamiltonian) and $H=H_0+V$ the scattering Hamiltonian. $V$
may be time dependent. We assume that $H$ and $H_0$ admit a good
scattering theory. In the time independent case, conservation of
energy is expressed as the statement that the scattering matrix
$S$ commutes with $H_0$ (not $H$!). Hence, for the frozen S-matrix
\begin{equation}
S_fe^{-iH_0 t}=e^{-iH_0t}S_f.
\end{equation}
This may be interpreted as the statement that the state $\psi$,
and the time shifted state $e^{-iH_0t}\psi$ both see the same
scatterer. Therefore it makes no difference if the time shift
takes place before or after the scattering.

This is, of course, not true in the time-dependent case. The
states $\psi$ and its time shift $e^{-iH_0t}\psi$ do not see the
same scatterer. Consequently,
the exact (dynamical) scattering matrix, $S_d$, for a time
dependent problem, acquires a time dependence:
\begin{equation}\label{eq:shift-Sd}
S_d(t)e^{-iH_0 t}=e^{-iH_0t}S_d.
\end{equation}
Now $S_d(t)$ will, in general, not coincide with $S_d$ except, of
course, at $t=0$ since energy is not anymore conserved. In the
time dependent case one does not have one scattering matrix, but
rather a family of them, $S_d(t)$. Since they are all related by
conjugation, any one of them is equivalent to any other. To pick
one, is to pick a reference point on the time axis. We shall use the
notation $S_d(0)=S_d$.

If $S_d$ is unitary\footnote{$S_d$ is unitary as a map between the spaces
of in and out states which may differ because states may get trapped or
released from the pump.}, which we henceforth assume, then ${\E }
_d=i\dot S_dS^*_d$ is Hermitian. We call it the ``energy shift"
for the following reason: $S_d$ satisfies the equation of motion
\begin{equation}
i\dot S_d(t)=[H_0,S_d(t)].
\end{equation}
Using the (assumed) unitarity of $S_d$, this can be reorganized as
\begin{equation}\label{energy-shift}
S_d(t)H_0S_d^*(t)=H_0-{\E } _d(t).
\end{equation}
Conjugation by the scattering matrix takes outgoing observables to
incoming observables. Eq.~(\ref{energy-shift}) justifies
identifying ${\E } _d$ with the operator of energy shift.
\begin{rem}
If we let $Q_j$ denote the projection on the states in the $j$-th
channel, then $\hat Q_j=S_dQ_jS^*_d$ projects on the out states
fed by the $j$-th channel. The energy shift generates the
evolution of $\hat Q_j$:
\begin{equation}\label{eq:dot-P-hat}
i\dot{\hat Q}_j= [\E_d ,\hat Q_j]
\end{equation}
\end{rem}

We are now ready to derive Eq.~(\ref{eq:rho-out}) which is an
operator identity for $\rho_{out}$. This is our starting point in
analyzing adiabatic transport. By the functional calculus we can
extend Eq.~(\ref{energy-shift}), evaluated at $t=0$, to
(measurable) functions of $H_0$, namely
\begin{equation}\label{eq:rho-out-2}
\rho_{out}=S_d\rho(H_0)S_d^*=\rho\big(H_0-{\E } _d\big).
\end{equation}
So far, no approximation has been made. The
identity does not assume an adiabatic time dependence.

\begin{rem} For comparison we establish the classical counterpart to
Eq.~(\ref{energy-shift}). Let $\Gamma$ be the classical phase space described
in Sect.~\ref{sec:cl-pumps} and let $\phi_s: \Gamma\to\Gamma$ be the
flow $\phi_s(E,t)=(E,t-s)$ generated by the Hamiltonian $h(E,t)=E$, i.e.,
the solution of the canonical equations of motion
\begin{equation}
\frac{d}{ds}\phi_s(x)=I(dh)\big|_{\phi_s(x)},
\end{equation}
where $x=(E,t)$ and $I: T^*\Gamma\to T\Gamma$ is the symplectic 2-form.
If times of passage are measured not w.r.t. time $0$ but w.r.t. time
$t$, then the scattering map, $S_d(t)$, satisfies,
cf. Eq.~(\ref{eq:shift-Sd}),
\begin{equation}\label{eq:shcl-Sd}
S_d(t)\circ\phi_t=\phi_t\circ S_d,
\end{equation}
where $S_d(0)=S_d$ has been introduced in Eq.~(\ref{eq:scl}). Since $S_d(t)$
is a family of symplectic maps its vector field is Hamiltonian:
\begin{equation}
\frac{d}{dt}S_d(t)(x)\big|_{t=0}=I(d\E_d)\big|_{S_d(x)},
\end{equation}
where $\E_d$ is a function on $\Gamma$ uniquely determined up to an additive
constant. By taking derivatives of (\ref{eq:shcl-Sd}) at $t=0$ we obtain
$I(d\E_d)+S_{d*}(I(dh))=I(dh)$. Since $S_d$ is symplectic we have $S_{d*}I=I$
and hence
$S_{d*}(I(dh))=(S_{d*}I)(S_{d*}dh)=I(S_d^*)^{-1}dh=Id(h\circ S_d^{-1})$, so
that we conclude
\begin{equation}
h\circ S_d^{-1}=h-\E_d,
\end{equation}
provided the constant not determined by $S_d$ is properly adjusted.
\end{rem}


\subsection{The Weyl calculus}
A convenient language for discussing the relation between operators in quantum
mechanics and functions on phase space, called symbols, is the Weyl calculus
\cite{ref:semiclassics,ref:robert}. For pumps the classical phase space
has been introduced at the beginning of Sect.~\ref{sec:cl-pumps}, with points
labelled by the pair
$(E,t)$, where $E$ is the energy of the (classical) particle and
$t$ its time of passage at the origin. An additional index $j=1,\ldots n$
labels the channels.

The relation of a (matrix valued) symbol $a(E,t)$ to the
corresponding operator $A$ is
\begin{equation}\label{eq:symbol-t}
\kernel{t,j}{A}{t',j'}=\frac 1 {2\pi}\int dE\, e^{-i(t-t')E}\
a_{jj'} \left(E,\frac{t+t'}2\right),
\end{equation}
where $\ket{t,j}$ is the (improper) state in the $j$-th channel whose
time of passage at the scatterer (in the $H_0$ dynamics) is $t$.
Equivalently,
\begin{equation}\label{eq:symbol-E}
\kernel{E,j}{A}{E',j'}=\frac 1 {2\pi}\int dt\, e^{i(E-E')t}\
a_{jj'} \left(\frac{E+E'}2,t\right),
\end{equation}
where $\ket{E,j}$ is the (improper) state in the $j$-the channel
with energy $E$. It follows that (if $A$ is trace class \cite{ref:simon})
\begin{equation}\label{eq:trace-symbol}
\Tr A =\frac 1 {2\pi}\int \tr a\, dE\,dt
\end{equation}
where $\tr a$ denotes a trace over channels, i.e. a trace of
finite dimensional matrices. Similarly, if $a$ or $b$ are
(locally supported) functions, we have
\begin{equation}\label{eq:prod-trace-symbol}
\Tr(AB) =\frac 1 {2\pi}\int \tr(ab)\, dE\,dt.
\end{equation}

\section{Adiabatic transport}\label{sec:ad-scattering}

The notion of approximation in adiabatic scattering requires some
explanation. In this regime the scattering of a particle occurs on the
time scale of the dwell time $\tau$ which is short compared
to the adiabatic time scale $\omega^{-1}$. Therefore,
the (unitary) operator $S_d$ should be related to
the frozen scattering matrices $S(E,t)$. While the uncertainty relation
forbids specifying both coordinates $E$ and $t$ of a particle, the variables
on which $S(E,t)$ actually depends are $E$ and $\omega t$.
This gives adiabatic scattering a semiclassical
flavor where $\omega$ plays the role of $\hbar$. Its theory can be phrased in
terms of the Weyl calculus
\cite{ref:semiclassics, ref:robert} with symbols which are
power\footnote{The dimensionless expansion parameter is $\omega\tau$.}
series in $\omega$. In particular, as we shall
explain, $S(E,t)$ may be interpreted as the principal symbol of
$S_d$. The chain of argument in making the identification goes
through the frozen S-matrix, $S_f(t)$, where the time of
freezing, $t$, is picked by the incoming state.

\subsection{Adiabatic scattering}

We shall show the following correspondence between operators and
symbols which, in our case, are $n\times n$ matrix functions of
$E$ and $t$:
\begin{eqnarray}
S_d(t_0)&\Longleftrightarrow &S(E,t+t_0)+O(\omega)\label{eq:s&symbol}\\
\E_d&\Longleftrightarrow &\E(E,t)+O(\omega^2)\label{eq:E&symbol}\\
\rho_{out}&\Longleftrightarrow &\rho(E)-\rho'(E)\big(\E(E,t)+
O(\omega^2)\big)+\half\rho^{\prime\prime}(E)\E^2(E,t)+O(\omega^3)\label{eq:rho&symbol}
\end{eqnarray}
Since the Fermi function at $T=0$ is a step function,
$\rho^\prime$ is a delta function and consequently, the notion of
smallness in the expansion in Eq.~(\ref{eq:rho&symbol}) is in the
sense of distributions\footnote{In agreement with
Eqs.~(\ref{eq:delta-rho}, \ref{eq:delta-rho-E}).}.

To see the first relation, let $\ket{t,j}$ denote the state that
traverses the scatterer at time $t$ and $S_f(s)$ denote the frozen
S-matrix associated with the Hamiltonian in effect at time $s$.
Then
\begin{equation}\label{eq:sf-sd}
\kernel{t,j}{S_d}{t',j'}=\kernel{t,j}{S_f\left(\frac{t+t'}2\right)}{t',j'}+O(\omega)
\end{equation}
The matrix elements on both sides are significant provided $t-t'$
is small, within the order of the dwell time, or the Wigner time
delay. Using
\begin{equation}\label{eq:fourier}
\ket{t,j}=(2\pi)^{-1/2}\int dE \,e^{iEt}\,\ket{E,j}
\end{equation}
one finds
\begin{equation}\label{eq:sf-to-s}
\kernel{t,j}{S_f\left(s\right)}{t',j'}=\frac 1 {2\pi}\int dE\,
e^{-i(t-t')E}\,S(E,s),
 \quad s=\frac{t+t'}2,
\end{equation}
and we have used the fact that $S_f$ is energy conserving. This
establishes Eq.~(\ref{eq:s&symbol}) for $t_0=0$ by comparison with
Eq.~(\ref{eq:symbol-t}). In the language of pseudo-differential
operators $S(E,t)$ is the principal symbol of $S_d$. More
generally, $S(E,t+t_0)$ is the principal symbol of $S_d(t_0)$. The
``quantization" of $S(E,t)$ then satisfies
Eq.~(\ref{eq:shift-Sd}), as it must.

Eqs.~(\ref{eq:E&symbol}, \ref{eq:rho&symbol}) now follow from the
rules of pseudo-differential calculus \cite{ref:robert}, and the
operator identity for the outgoing states Eq.~(\ref{eq:rho-out}).

\subsection{Currents}\label{sec:BPT}
A rigorous derivation of BPT is presented in \cite{ref:aegss,ref:schnee}.
Here, instead, we shall be content with a formal, but relatively
straightforward derivation using Weyl calculus.

Let $Q_j^{in/out}(x,t_0)$ be the observable associated with
counting the incoming/outgoing particles in a box that lies to the
right of a point $x$ in the $j$-th channel at the point in time
$t_0$. The point $x$ is chosen far from the scatterer, but not so
far that the time delay relative to the pump is of order
$\omega^{-1}$. Namely, $v\tau\ll x\ll v/\omega$. The symbol of
$Q^{in/out}$ is a matrix valued step function:
\begin{equation}\label{eq:F-switch}
Q_j^{in/out}(x,t_0)\Longleftrightarrow P_j\,\theta\big(
v(t_0-t)-x\big)\theta(\mp(t_0-t)),\quad
\left(P_j\right)_{ik}=\delta_{jk}\delta_{ij}, \quad
v=\epsilon^\prime(p),
\end{equation}
and $P_j$ is the projection matrix on the $j$-th channel. Indeed, the
position of a particle with coordinates $(E,t)$ at time $0$ will be
$-v(t-t_0)$ at time $t_0$, see Eq.~(\ref{eq:e-t-canonical}). The particle
will then be outgoing if $t-t_0<0$.
Notice that $t=t_0$ falls outside of the support of the first Heaviside
function. The associated incoming/outgoing current operators are
\begin{equation}\label{eq:current-operator}
{\dot
Q}_j^{in/out}(x,t_0)=i[H,Q_j^{in/out}(x,t_0)]=i[H_0,Q_j^{in/out}(x,t_0)].
\end{equation}
Here we used the fact that beyond $x$, deep inside the channel,
$H(t)$ coincides with $H_0$. The symbol associated with the
current is most easily computed recalling that in Weyl calculus
commutators are replaced by Poisson brackets. This reproduces the
usual notion of a current
\begin{equation}\label{eq:qdot-symbol}
\dot Q_j^{in/out}(x,t_0)\Longleftrightarrow
P_j\,\{E,\theta\big(v(t-t_0) -x\big)\}= \mp P_j\,
\delta\big(t-t_0-x/v\big)\theta(\mp(t_0-t)),
\end{equation}
$x$ is where the ``ammeter'' is localized. By the assumption that
the ammeter is not too far it leads to a slight modification of
$t_0$, the time when current is measured. We henceforth drop $x$.
Now the expectation value of the current is
\begin{equation}
\average{\dot Q}_j(x,t_0)=\Tr \big(\rho_{out} Q_j^{out}\big)+\Tr
\big(\rho Q_j^{in}\big)=\Tr\big(\delta\rho \ \dot
Q_j^{out}(x,t_0)\big),\quad \delta\rho=\rho_{out}-\rho.
\end{equation}
Using Eq.~(\ref{eq:prod-trace-symbol}) to evaluate the trace we
find
\begin{eqnarray}\label{eq:bpt-symbols}
\average{\dot Q}_j(x,t_0)&=&-\frac 1 {2\pi}\int
dE\,dt\,\rho^\prime(E)
\E_{jj}(E,t)\delta(t-t_0)+O(\omega^2)\nonumber \\ &=&-\frac 1
{2\pi}\int dE\,\rho^\prime(E) \E_{jj}(E,t_0)+O(\omega^2)
\end{eqnarray}
 reproducing Eq.~(\ref{eq:BPT}).


\subsection{Dissipation}\label{sec:dissipation}
To compute the dissipation we start as in the previous section.
Let $D_j^{in/out}$ denote the observable associated with the
incoming/outgoing excess energy in the $j$-th channel in a box to
the right of the point $x$. The excess energy is, of course,
energy measured relative to the Fermi energy:
\begin{equation}\label{eq:dissipation-observable}
D_j^{in/out}(x,t_0)= \half\{Q_j^{in/out}(x,t_0),H_0-\mu\}
\end{equation} and $Q_j^{in/out}$ is as in Eq.~(\ref{eq:F-switch}). The observable
associated with dissipation current in the $j$-th channel is the
time derivative of the excess energy, i.e.,
\begin{equation}
\dot D_j^{in/out}(x,t_0)=\dot E_j^{in/out}(x,t_0)-\mu\dot
Q_j^{in/out}(x,t_0) =i\left[H_0,D_j^{in/out}(x,t_0)\right].
\end{equation}
The symbol corresponding to the dissipation current is then
\begin{equation}\label{eq:dissipation-symbol}
D_j^{in/out}(x,t_0)\Longleftrightarrow
\mp P_j\,(E-\mu)\,\delta\big(t-t_0-x/v\big)\theta(\mp(t_0-t)).
\end{equation}
The expectation value of the dissipation current is therefore
\begin{equation}\label{eq:dissipation-def}
\average{\dot D}_j(x,t_0)= \Tr\big(\delta\rho\,\dot
D_j^{out}(x,t_0)\big).
\end{equation}
We shall now show that for $T\lesssim\sqrt{\omega/\tau}$ the dissipation
is quadratic in $\omega$ and is determined by
Eq.~(\ref{eq:dissipation}).

As in section~\ref{sec:BPT} we shall evaluate the trace using
Eq.~(\ref{eq:prod-trace-symbol}). At low temperature $\rho^\prime$
is concentrated near the Fermi energy. We may then approximate the
energy shift up to its linear variation near $\mu$. For the
term proportional to $\rho^\prime$ in the expansion
Eq.~(\ref{eq:rho&symbol}) the contribution to the dissipation is
proportional to
\begin{eqnarray}\label{eq:rhoprime-dissipation}
-\frac 1 {2\pi}\int
dE\,\rho^\prime(E)\left(\E_{jj}(\mu,t_0)+(E-\mu)\E^\prime_{jj}(\mu,t_0)+O(\omega^2)\right)
 (E-\mu)\nonumber \\
 =O\left(\beta
e^{-\beta\mu}\right)+O(\omega T^2)+O(\omega^2 T).
\end{eqnarray} The term proportional to
$\rho^{\prime\prime}$ in the expansion gives,
\begin{equation}\label{eq:rho-dprime-dissiparion}
\frac 1 {4\pi}\int dE\,
\rho^{\prime\prime}(E)(\E^2)_{jj}(E,t_0)(E-\mu)=\frac 1
{4\pi}(\E^2)_{jj}(\mu,t_0)+O(\omega^2 T),
\end{equation} which is the requisite result,
Eq.~(\ref{eq:dissipation}).

The result (\ref{eq:dissipation}) is remarkable in that we obtain
the dissipation to order $\omega^2$ by making two approximations,
each valid only to order $\omega$. First we replace $\E_d$ with
$\E$, and then we evaluate $\E$ at $E=\mu$. Had we used this
procedure to compute the quantities $\dot E_j$ and $\dot Q_j$
separately, each of them would be off by nonzero $O(\omega^2)$
terms (as can be seen in the snowplow and battery examples);
nevertheless, the combination $\dot E_j - \mu \dot Q_j$ is
computed correctly to order $\omega^2$. The reason is that in
each quantum channel one has the following lower bound on the
dissipation \cite{ref:AEGS:PRL}:
\begin{equation}\label{eq:optimal}
\dot E_j - \mu \dot Q_j\ge \pi\dot Q^2_j.
\end{equation}
This bound is saturated by the outgoing population
distribution that is filled up to energy $\mu$ and empty
thereafter. The dissipation should therefore be quadratic in the
deviation of the outgoing distribution from this minimizer. Since
the outgoing distribution is an $O(\omega)$ perturbation of the
minimizer, knowing the distribution to order $\omega$ should give
the dissipation to order $\omega^2$.

\begin{rem}
There would appear to be
two problems with the argument above. In minimizing a functional
on a region with a boundary, one obtains a quadratic
estimate for the functional around its minimizer if the minimum
occurs at an interior point. If the minimum occurs at the boundary, then
a variation away from the boundary can increase the function to first order.
Furthermore, whether the minimum occurs at an interior point or on the
boundary, quadratic estimates depend on the Hessian being a bounded operator.
If the Hessian is unbounded, then an arbitrarily small change in the point
can cause an arbitrarily large increase in the functional. In our case,
the minimum occurs at a point that is on the boundary of the constraints
$0 \le \rho(E) \le 1$. There are large modes for the Hessian, involving
adding electrons at arbitrarily high energies.

Fortunately, neither exception is relevant. In fact, the
correction of the distribution, as given by (\ref{eq:rho&symbol}),
consists of a local reshuffling of electrons around the Fermi
energy and does not involve the large modes of the Hessian. These
variations should not be viewed as being either towards or away
from the boundary, since neither the expression
(\ref{eq:rho-out-2}) nor its opposite (replacing $\E$ with
$-\E$) violates the constraints.
\end{rem}
\subsection{Entropy and noise currents}
\label{sec:entropy-noise} Entropy and noise introduce a new
element in that the transport equation,
Eq.~(\ref{eq:entropy-current}), depends on the density through a
{\em non-linear} function $h(\rho)$. For the entropy and noise
$h(x)$ is given in Eq.~(\ref{eq:h-def}). Using the fact that in
either case $h(0)=h(1)=0$, we shall
show that for $\omega\ll T\ll \sqrt{\omega/\tau}$ the currents are
quadratic in $\omega$ and are given by
\begin{equation}\label{eq:entropy-2-order-final}
\dot s_j(t,\mu,T)=\frac \beta {2\pi }
\,\Delta\E_j^2\,(\mu,t)\big)\int_0^1 dx \,h(x)\:,
\end{equation}
where $\Delta\E_j^2$ has been defined in
Eq.~(\ref{eq:delta-E-2}). For the entropy the integral gives $1/2$
and for the noise it gives $1/6$. To complete the derivation of
the noise and entropy currents, we now derive
Eq.~(\ref{eq:entropy-2-order-final}).

The condition $\omega\ll T$ makes it possible to consider the outgoing state
of the electrons with a fixed time of passage, provided the time resolution is
short compared to $\omega^{-1}$ but large w.r.t. $T^{-1}$. The state
$\rho_{out,j}=P_j\rho_{out}P_j$ is then given, see (\ref{eq:rho&symbol}), as
\begin{equation}
\rho_{out,j}(E)=\rho(E-{\E }_{jj}(E,t))+
\frac{1}{2}\rho^{\prime\prime}(E)
\bigl(({\E } ^2)_{jj}(E,t)-{\E }_{jj}(E,t)^2\bigr).
\end{equation}
The entropy/noise current (\ref{eq:entropy-current}) is
\begin{equation}
\frac {1}{2\pi }\int dE\,\bigl(
(h\circ \rho)(E-{\E }_{jj}(E,t))-(h\circ \rho)(E)\bigr)
+\frac {1}{4\pi }\int dE\,(h^\prime\circ \rho)(E)\rho^{\prime\prime}(E)
\Delta\E_j^2\,(E,t).
\end{equation}
In these integrals, ${\E }_{jj}$ may be regarded as constant in $E$
because of the condition $T\ll \sqrt{\omega/\tau}$. The first integral then
vanishes and in the second we may pull $\Delta\E_j^2\,(\mu,t)$ out of
the integral. This leaves us with the integral
\begin{equation}\label{eq:entropy-2-order-2}
 \int dE\, (h^\prime\circ \rho)(E)\,
 \rho^{\prime\prime}(E)= -\beta\int_0^1 d\rho\, h^\prime(\rho)\,
 (1-2\rho)= 2\beta\int_0^1 d\rho\, h(\rho),
\end{equation}
where, in the second step, we have used a property of the Fermi function,
$\rho^\prime=-\beta\,\rho(1-\rho)$, and an integration by parts in the
last step. This establishes the result.

We have nothing to say about the range $T\lesssim\omega$. The noise at
$T=0$ can be calculated using a formalism of Lesovik and Levitov that we
discuss in the appendix. This formula is nonlocal in time. It is
instructive to examine what goes wrong with our approach at $T=0$.
In this limit $\rho'$ and $\rho^{\prime\prime}$ are distributions
and $\rho$ a discontinuous function. Since it is not allowed to
multiply distributions by distributions, or even by discontinuous
functions, equations such as Eq.~(\ref{eq:entropy-2-order-2})
which are not linear in $\rho$ make no sense. This is a
reflection of the fact that in the regime where $T\lesssim\omega$ entropy
and noise currents have a memory that goes back in times of order
$\beta$. In the regime we consider $\omega\ll T$ the memory is
short compared with the time scale of the pump and instantaneous
formulas make sense. At the opposite regime, where $T\lesssim\omega$,
a local formula in time cannot be expected.
\section{Examples}\label{se:exa}

Quantum pumps may be viewed either as particle pumps or as wave
pumps. The particle interpretation has a classical flavor where
the driving mechanism is identified with forces on the particles.
The wave interpretation stresses the role played by phases and
suggests that interference phenomena play a role. This duality can
be seen in the BPT formula in the two channel case of section
\ref{sec:bpt}. On the one hand Eq.~(\ref{dq2}) makes it clear that
the phases in the $S$-matrix, $(\alpha,\phi,\gamma)$, play a role
in charge transport. In fact, changing the transmission and
reflection probabilities while keeping the phases fixed cannot
drive a current. At the same time, the rate of change of two of
the three phases, $\dot\phi$ and $\dot\alpha$, also admit a
classical interpretation as EMF and Galilean shift. The pedestrian
derivation is a reflection of the fact that particle
interpretation is more intuitive. Here we shall consider two
examples where dual reasoning is insightful.

\subsection{The bicycle pump}
\label{se:bikepump}

\begin{figure}[h] \hskip 2. cm
\includegraphics[width=10.cm]{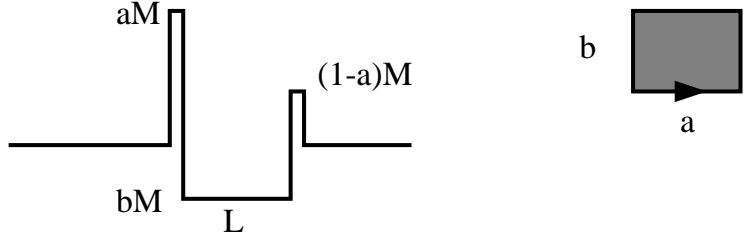} \caption{On the left a
typical configuration of the potential of the quantum bicycle pump. On
the right a loop in the $a$-$b$ parameter space.}\label{barrieres:fig}
\end{figure}

In a bicycle pump the action of the valves is synchronized with
the motion of the piston. The analogous quantum pump has
synchronized gates as shown in Fig.~\ref{barrieres:fig}. The
particle interpretation of the pump is simple and intuitive. The
wave (or BPT) point of view is more subtle. In particular as we
shall see, in terms of the elementary processes described in
section \ref{sec:bpt} the pump operates by changing the phases
$\gamma$ and $\alpha$: Galilean shifts arise from the synchronized
action of the gates.

Let us choose a length scale so that $k_F=\pi$ and an energy scale so
that $\mu=1$. In these units, choose the length of the pump, $L$, to be
an integer $L=n$, pick the valves thin, $\delta \ll 1$, and
impenetrable, i.e., of height $M$ with $M\delta\gg 1$. Consider the
potential, shown in Fig.~\ref{barrieres:fig}, that depends on two
parameters, $a$ and $b$, that vary on the boundary of the unit
square $[0,1]\times[0,1]$:
\begin{equation}
\label{bikepot}
V_{a,b}(x)  =
  \begin{cases}
    0 & \text{if $x<0$,}, \\
   a M  & \text{if $0 \le x < \delta$,}\\
   10\,b  & \text{if $\delta \le x < L$}\\
   (1-a)M & \text{if $L \le x < L+\delta$}\\
    0 & \text{if $x \ge L+\delta$.}
  \end{cases}
\end{equation}
Since the quantum box was designed to accommodate $n$ particles,
the pumps transfers $n$ particles in each cycle. Like in the
bicycle pump, at all times, at least one of the valves is closed.
This give the particle point of view \footnote{For related results
see e.g. \cite{ref:entin}.}.

We now consider this pump from the perspective of the phases in
BPT. At all stages of the cycle, the transmission coefficients at
the Fermi energy are essentially zero, so, by unitarity, the
reflection coefficients $\mathrm{r}$ and $\mathrm{r}'$ are unit
complex numbers. The phases of the reflection amplitude,
$\gamma\pm\alpha$ of Eq.(\ref{S1}), must therefore change by $\pm
2\pi n$ in each cycle of the pump. How does this happen? As we
shall see in spite of the fact that the pump is operated by
manipulating gate voltages, the interpretation in terms of the
$S$-matrix is in terms of an interplay between Galilean shifts
$d\alpha$ and the Birman-Krein term $d\gamma$.

At $a=0, b=1$, we have $\mathrm{r}=-1$, since the piston imposes a
Dirichlet condition at $x=0$, and $\mathrm{r}'=- \exp(2ik_F L)=-1$,
since the valve on the right imposes a Dirichlet condition at
$x=L$. As $b$ is decreased, $\mathrm{r}'$ remains $-1$ (since
the right valve is closed) and by Eq.~(\ref{S1})
$d\alpha=d\gamma$. Meanwhile, the wave functions of incoming waves
from the left penetrate deeper and deeper into the region $0<x<L$,
eventually accumulating $n$ half-wavelengths. The phase of
$\mathrm{r}$ increases by $2 \pi n$, since the left barrier has
been effectively shifted a distance $L$ to the right and $\int
dQ_1  = -n$.

The path with $b=0$ has no effect on the scattering matrix, since
$L$ is an integral number of half-wavelengths, so a Dirichlet
condition at $x=0$ is equivalent to a Dirichlet condition at
$x=L$. The left valve closes and the right valve opens but no
current flows. The remaining two legs of the path can be similarly
analyzed. Increasing $b$ with $a=1$ decreases the phase of $\mathrm{r}'$
by $2 \pi n$,
while decreasing $a$ with $b=1$ has no effect on $\mathrm{r}$ or
$\mathrm{r}'$.

The fact that $\mathrm{t}=\mathrm{t}'=0$ throughout the process
might seem strange. After all, how can you transport particles
without transmission? However, this is exactly what happens with
macroscopic pumps. Good bicycle pumps are typically transmissionless,
while bad pumps have leaky valves.

\subsection{The U-turn pump}\label{se:Uturn}
\begin{figure}[h]
\hskip1.5 in\includegraphics[width=8cm]{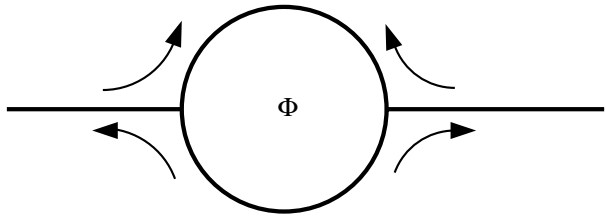}
 \caption{The graph associated with a model of the quantum Hall
effect.}\label{fig:qhe}
\end{figure}

The U-turn pump, shown in Fig. \ref{fig:qhe}, is a highly schematic
version of the quantum Hall pump.
There are two leads connected to a loop of
circumference $\ell$. The loop is threaded by a slowly varying
flux tube carrying a flux $\Phi$. The particle satisfies the free
Schr\"odinger equation on the edges of the graph and and satisfies
an appropriate boundary condition at the vertices. The boundary
conditions are such that at the Fermi energy all particles are
forced to make a U-turn at the loop. Namely, all particles coming
from the right circle the loop counter-clock\-wise and the exit on
the right while all those coming from the left circle the loop
clockwise and exit on the left.

Let us first look at this pump as a wave pump.
Since all particles make a U-turn, the transmission amplitude
vanish and the reflection amplitudes are phases. A left mover on
the loop accumulates a Bohm-Aharonov phase in addition to the phase
due to the ``optical length'' of the path. This means that the
$S$-matrix at the Fermi energy is
\begin{equation}
S(\mu,\Phi)=\begin{pmatrix}
  e^{i(k_F\ell+\Phi)} &0\\
0 &e^{i(k_F\ell-\Phi)}
\end{pmatrix},
\label{eq:hall-scatt}
\end{equation}
By BPT the charge transport is
\begin{equation}\label{eq:hall-pump}
  \average{dQ}_\pm(\mu)= \pm\frac{d\Phi}{2\pi}
\end{equation}
One charge is pumped from left to right in a cycle of the pump as
$\Phi$ increases by $2\pi $, the unit of quantum flux.

The scattering calculation, although easy, does not really explain
how the pump operates: How does it transport charges from right to
left if all charges are forced to make a U-turn at the loop? The
particle interpretation demystifies the pump: Particles in the
loop see a force associated with the EMF $\dot\Phi$ which make the
clockwise movers feel as if they are going uphill while the
counter-clockwise movers all go downhill. Because of this some of the slow
counter-clockwise movers turn into clockwise movers and exit on the
other side. Although insightful, the particle interpretation does
not readily translate to a qualitative computation without
invoking some wave aspects.

\begin{figure}[hall]
 \centerline{\epsfxsize=3truein
 \epsfbox{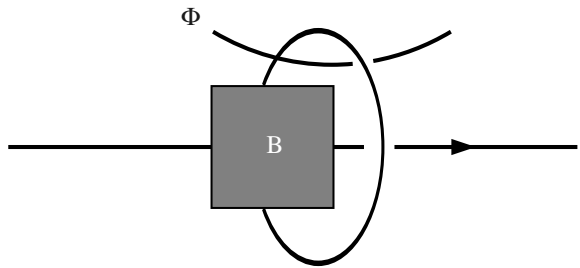}
 }
\caption{The Hall effect as a pump driven by the EMF $\dot \Phi$.}
\label{qhall}
\end{figure}

The integer quantum Hall effect \cite{dolgo} can
be described by a scatterer with four
leads (north, south, east and west) with a north-south voltage and an
east-west current. However, if the north and south leads are
connected by a wire, and if the resulting loop is threaded by a
time-varying magnetic flux to generate the north-south voltage, then
one obtains a geometry shown in Fig.~{\ref{qhall}}.

The U-turn pump of Fig.~{\ref{fig:qhe}} models the essential
features of this geometry. The 2D electron gas in
the Hall sample, and the magnetic field applied to the Hall
crystal, are modelled by the vertices, which scatter particles in
a time-asymmetric manner. The clockwise movers of
Fig.~{\ref{fig:qhe}} correspond to electrons that enter the Hall
bar from the west, move along the edge of the crystal until they
reach the south lead, go along the loop from south to north, move
along the edge from north to west, and emerge to the west. The
counter-clockwise movers correspond to electrons that go from east
to north (along the Hall bar) to south (along the loop) and then
to east and out the east lead. By standard arguments \cite{fgw},
the edge states reflect the existence of localized bulk states in
the crystal.

\subsection{A family of optimal pumps}

Optimal pumps \cite{ref:AEGS:PRL} saturate the bound in
Eq.~(\ref{eq:optimal}). In \cite{ref:alekseev} it was shown that
optimal pumps that do not break time-reversal are
transmissionless. (The two examples above also fall into the
category of being optimal and transmissionless.) The following
example shows that a general optimal pump can have any value of
$\mathrm{r}$ and $\mathrm{t}$.

In the battery of section \ref{subse:bat} the scatterer got in the
way of the electrons, and the most efficient transport was with
$\mathrm{r}=0$. In the snowplow of section \ref{subse:snowplow}
the scatterer pushed the electrons and the most efficient
transport was with $|\mathrm{r}|=1$. In the following example of
optimal pump we combine a voltage with a moving scatterer, such
that the scatterer is moving along with the electrons, neither
pushing nor getting in the way. In this case, the scatterer
doesn't actually {\it do} anything, and we get efficient
transport, regardless of the initial values of $(\mathrm{r, t, r', t'})$.

Write the scattering matrix of a system where $\alpha=2\mu \xi$
and $\phi $ evolve as
\begin{equation} S(\mu) = \begin{pmatrix}
 \mathrm{r} e^{-2 i \mu \xi} & \mathrm{t}' e^{-i \phi }  \\
\mathrm{t} e^{ i \phi } &\mathrm{r}' e^{2 i \mu \xi}
\end{pmatrix}, \end{equation}
Synchronizing the velocity $\dot\xi$ with the voltage $\dot\phi$
according to
\begin{equation}
2\mu \dot\xi=\dot\phi\
\end{equation}
makes $\dot S=i\dot\phi\sigma_3 S$. The energy shift is then a
diagonal matrix
\begin{equation} {\E }  = i \dot S S^* =
-\dot\phi\ \sigma_3\end{equation} which implies that the pump is
optimal.

\subsection{The phase space of a snowplow}\label{sec:phasespace}

We give a description of a classical snowplow moving on the real axis at speed
$v_0$ during the time interval $[-T,T]$, but at rest before and after that.
It is described by the (total) phase space $\mathbb{R}^2\ni(x,p)$ with
Hamiltonian function $h=p^2/2+V(t)$, where $V(t)$ is a barrier of fixed
height $V>v_0^2/2$ and zero width located at
\begin{equation}
x(t)=\begin{cases} -v_0T,\quad&(t\le -T),\\
v_0t,&(-T<t<T),\\
v_0T,&(t\ge -T).
\end{cases}
\end{equation}
We use the notation of Sect.~\ref{sec:cl-pumps} and denote by 1 the left
channel ($x<0$) and by 2 the right one ($x>0$).
Let $\Gamma_{ij}^+\subset\Gamma_j\subset \Gamma^+$ be the outgoing labeled
trajectories originating from channel $i$ and eventually ending in channel
$j$. The same meaning has $\Gamma_{ij}^-\subset\Gamma_i\subset \Gamma^-$,
except that the trajectories are incoming labeled.

\begin{figure}[h]
\begin{center}
\input{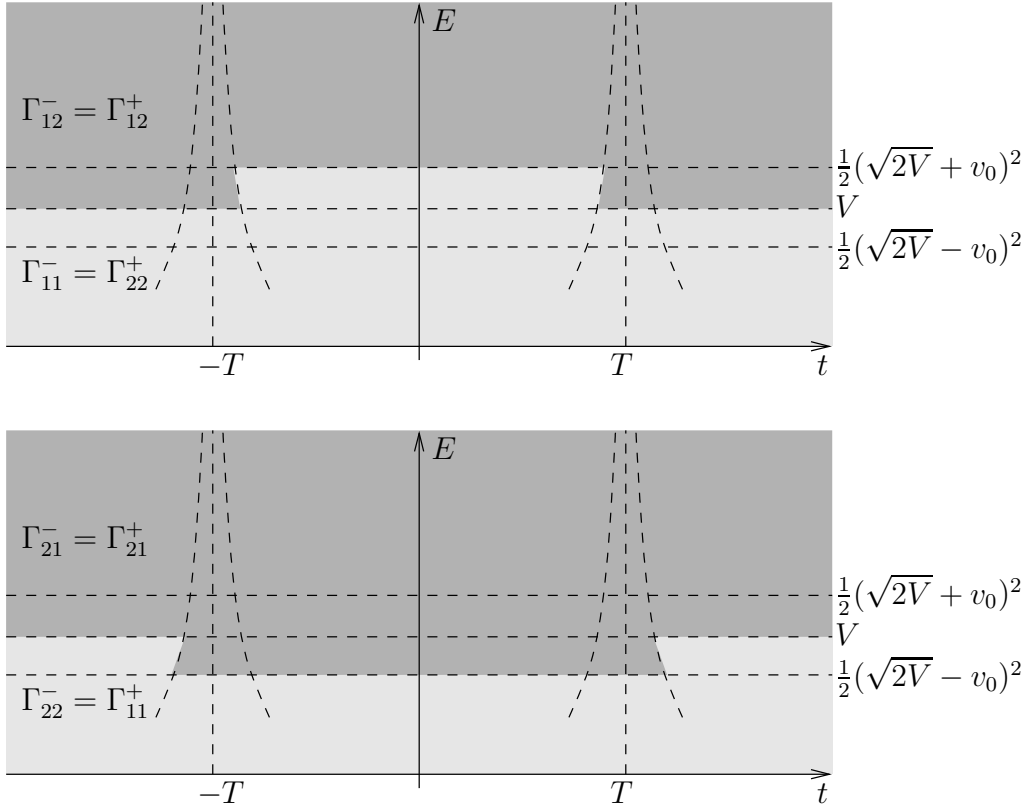}
\caption{Portions of phase space corresponding to transmitted (dark
shaded) and reflected (light shaded) trajectories with labels $+/-$
corresponding to outgoing/incoming data. The curves on the right
halves correspond to $t=T\pm\frac{v_0}{\sqrt{2E}}T$.}
\label{fig:phasespace}
\end{center}
\end{figure}

If a particle crosses the above scatterer of zero width, then its
trajectory is free at all times. Hence
\begin{equation}
\Gamma_{12}^-=\Gamma_{12}^+,\quad \Gamma_{21}^-=\Gamma_{21}^+.
\end{equation}
It suffices to compute these subsets only. The remaining ones are then given
by complementarity:
\begin{equation}
\Gamma_{i1}^-\cup\Gamma_{i2}^-=\Gamma_i,\quad
\Gamma_{1j}^+\cup\Gamma_{2j}^+=\Gamma_j,
\end{equation}
with disjoint unions. \\
\begin{itemize}
\begin{item}{$\Gamma_{12}^-=\Gamma_{12}^+$.}
\begin{figure}[h]
\begin{center}
\input{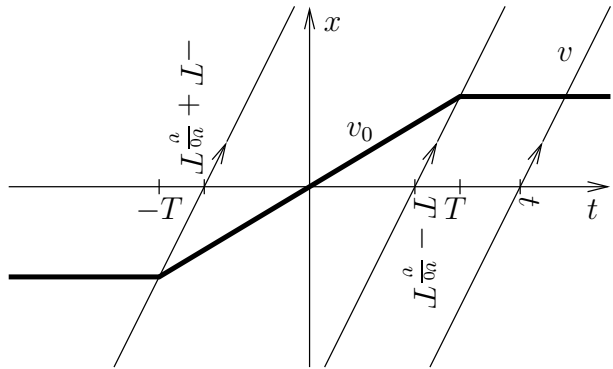}
 \caption{The thick line represents the position $x(t)$ of the snowplow.
The other lines are free trajectories of common energy $E$ indicated by the
slope $v=\sqrt{2E}>0$. Their times of passage $t$ are read off at the
intercepts with the abscissa.}\label{fig:snowplow}
\end{center}
\end{figure}
Depending on their time of passage $t$, trajectories of this type will
(see Fig.~\ref{fig:snowplow}) require an energy
$E=v^2/2>E_c\equiv v_c^2/2$, with critical energy $E_c$ given as
\begin{equation*}
E_c=V\quad\hbox{if}\quad |t|>T-\frac{v_0}{v}T=T-\frac{v_0}{\sqrt{2E}}T;
\end{equation*}
resp. by $v_c-v_0=\sqrt{2V}$, i.e.,
\begin{equation*}
E_c=\frac{1}{2}(\sqrt{2V}+v_0)^2\quad\hbox{if}\quad
|t|\le T-\frac{v_0}{\sqrt{2E}}T.
\end{equation*}
This portion of phase space is drawn dark shaded in the upper part
of Fig.~\ref{fig:phasespace}.
\end{item}
\begin{item}{$\Gamma_{21}^-=\Gamma_{21}^+$.} In this case the slope of
free trajectories is $v=-\sqrt{2E}<0$. As a result the critical energy
$E_c$ is
\begin{eqnarray*}
E_c=V\quad&\hbox{if}&\quad |t|>T+\frac{v_0}{\sqrt{2E}}T,\\
E_c=\frac{1}{2}(\sqrt{2V}-v_0)^2\quad&\hbox{if}&\quad
|t|\le T+\frac{v_0}{\sqrt{2E}}T.
\end{eqnarray*}
This portion of phase space is drawn dark shaded in the lower part of
Fig.~\ref{fig:phasespace}.
\end{item}
\end{itemize}

\subsection{Classical scattering from a battery}\label{sec:nophase}

This example  shows that, in the classical case, static scattering
data {\it cannot} determine the energy shift. Consider the
classical version of the battery, see Sect.~\ref{subse:bat}, with
Hamiltonian function $h(x,p)=(p-A)^2/2$ and gauge
$A(x,t)=t\phi'(x)$ of compact support. Clearly, this describes
particles which get accelerated as they cross the pump, whence
there is an energy shift. More formally, the quantity
$f(x,p)=(p-A)^2/2+\phi$ is a constant of motion, as it is verified
from Newton's equation $(d/dt)(p-A)=-\dot A=-\phi'$. In the leads,
$f=p^2/2+\phi$, which implies that the energy $E=p^2/2$, as
defined there, gets shifted by $\E=-\phi|^\infty_{-\infty}$ if a
particle crosses the battery from left to right. On the other
hand, for each static scatterer, $p-A$ is a constant of motion,
whence the static scattering maps all equal the identity map! The
quantum mechanical phase information, which was present in the
static $S$-matrix and determined the energy shift, is of course
unavailable here.

\section{Geometry and topology}\label{sec:geometry-topology}

When a pump goes through a cycle,
so does the scattering matrix $S(E,t)$. The rows of matrix define
unit vectors in $\mathbb{C}^n$ and the charge transport in the
$j$-th channel can be interpreted in terms of geometric properties
of these vectors.

As we shall explain in section \ref{se:globalangle}, the charge
transport is naturally identified with the global angle
accumulated by a row of the S-matrix during a cycle. {\em It need
not be an integer multiple of $2\pi$}. Computing this angle is
formally the same as computing Berry's phase \cite{ref:berry}.
That the global phase has direct physical significance is related
to the fact that a quantum pump is a wave pump.

Interestingly, the charge transport {\em in a closed cycle} can
also be computed by forgetting altogether about the (global) phase
provided one has knowledge about what happens on a surface
spanning this cycle. This reflects a basic geometric relation
between the failure of parallel transport and the curvature.

\subsection{Charge transport and Berry's phase}

Let
\begin{equation}
\label{S-ket}
\ket{\psi_j} =
\begin{pmatrix}
S_{j1} \\ S_{j2} \\  \vdots \\ S_{jn}
\end{pmatrix} \in \mathbb{C}^n \end{equation} be
the transpose of the $j$-th row of the frozen $S$ matrix,
evaluated at the Fermi energy. Yet another rewriting of the BPT
formula at zero temperature is as
\begin{equation} 2 \pi \average{dQ}_j = i \langle
\psi_j | d \psi_j \rangle . \label{eq:dQbra-ket} \end{equation}
The expression on the right hand side is familiar from the context of adiabatic
connections and Berry's phase \cite{ref:berry}.
To simplify notation we shall fix a row, say the first, and drop the index $j$.

There are important conceptual differences between the Berry's
phase associated to a quantum state $\ket{\psi}$ and the phases
that arise in the study of a row of the S-matrix $\ket{\psi}$.
In the usual Berry's phase setting one starts with a circle of
Hamiltonians to which one associates a (unique) circle of {\em
projections} $\vert\psi\rangle\langle\psi\vert$, say on the ground
state. To represent these projections in terms of a circle of
eigenvectors $\ket{\psi}$ one needs to choose a reference phase
arbitrarily for each point on the circle. The physical adiabatic
evolution picks its own phase relative to the reference phase.
Berry's phase then measures the phase accumulated in a
cycle\footnote{By choosing the energy of the state to be zero one
can always get rid of the dynamic phase.}. In particular, it does
not make sense to talk about the phase accumulated on a path that
is not closed.

In contrast, a cycle of the pump is a circle of vectors
$\ket\psi$ of the scattering matrix (not a cycle of
projections). No choice of a phase needs to be made---the phases
are all fixed by solving the scattering problem. In particular,
the vector $\ket{\psi}$ always returns to itself after a complete
cycle of the pump. Moreover, unlike the usual Berry's phase,
Eq.~(\ref{eq:dQbra-ket}) makes perfectly good physical and
mathematical sense also for an open path and not just for a closed
cycle.

What, then, does the Berry's phase measure in the present context?
This is addressed in the next section.
\subsection{Global angle}\label{se:globalangle}
As we now explain the charge transport in a quantum pump has a
geometric interpretation in terms of the global angle of
$\ket{\psi}$.

The notion of global angle is obvious in the case that the circle
of vectors are all parallel, i.e.,
\begin{equation}
\ket{\psi}=e^{i\gamma}\ket{\psi_0}
\label{eq:grcirc}
\end{equation}
with $\ket{\psi_0}$ fixed. The global angle is $\gamma$ and is related
to the charge $2\pi\average{ dQ}=-d\gamma$.

Now, how do we compare global angles when vectors are not
parallel? This is a classical problem in geometry whose answer
relies on the notion of parallel transport. Namely, one needs a
rule for taking one vector to another without changing its global
phase. A natural way to do so is to impose that there be no motion
in the ``direction'' $|\psi\rangle\langle\psi |$. Explicitly, we
say that $\ket{\psi}$ is parallel transported if
\begin{equation}
|\psi\rangle\langle\psi |d\psi\rangle=0
\end{equation}
We can now compare the global phase for any two vectors connected
by a path. The path may be either open or closed, and the global
angle makes sense in either case. The right hand side of
Eq.~(\ref{eq:dQbra-ket}) is (minus) the change in the global
angle and the left hand side identifies it with the transported
charge. This identifies charge transport with a global angle.

The global angle need not change by a multiple of $2\pi $ in a
cycle of the pump---except in the special case that
$\ket{\psi}=e^{i\gamma}\ket{\psi_0}$ with $\ket{\psi_0}$ fixed.
The failure of parallel transport for closed paths is interpreted
in geometry as curvature.
\subsection{Curvature}\label{sec:curvature}
In the previous section we identified charge transport with the
global angle of $\ket{\psi_j}$. Remarkably, one can compute the
charge transport {\em in a closed cycle} while forgetting
altogether about the global phase and relying only on the
projection $\ket{\psi_j}\bra{\psi_j}$. For a closed cycle one can
relate the line integral on the boundary of a disk $\partial D$ to
a surface integral on the disk $D$ via Stokes
\begin{equation}\label{eq:q-curvature}
2\pi\average{Q}_j= i\int_{\partial D} \bra{\psi_j}d\psi_j\rangle=
i\int_D \bra{d\psi_j}d\psi_j\rangle
\end{equation}
In the context of pumps the identity is known as Brouwer's formula
\cite{ref:Brouwer}.

\begin{rem}
For other ways to rewrite Eq.~(\ref{eq:q-curvature}), use the identities
\begin{equation}
i \bra{d\psi_j}d\psi_j\rangle=-i(dS\wedge dS^*)_{jj}=
-i\Tr\hat P_j(d\hat P_j\wedge d\hat P_j)\hat P_j,
\label{eq:p-curvature}
\end{equation}
where $\hat P_j=S^*P_jS$ is the projection onto the state
feeding channel $j$. The r.h.s. is the trace of the
curvature of the connection $\hat P_jd$ (see e.g. \cite{ref:avron}, Sect. 9.5).
\end{rem}

We shall now describe a different
interpretation of Brouwer's formula that focuses on the Wigner time
delay and the energy shift. We consider the charge transport in a
cycle, so $t$ is an angle. By Eq.~(\ref{pcf}), and assuming
no semi-bound states, so $\E(0,t)=0$, the charge transport in a
cycle is
\begin{equation}\label{eq:QasTE-uncty}
2\pi \average{Q}_j= \int_C dE\wedge dt \,\Omega_{jj}
\end{equation}
where $C$ is the cylinder $[0,\mu]\times S$. Now
\begin{equation}
\Omega_{jj}\,dE\wedge dt=d\E_{jj}\wedge dt+dE\wedge d{\cal T}_{jj}
=-i\,(dS\wedge dS^*)_{jj}.
\end{equation}
The difference between this formula and Eq.~(\ref{eq:q-curvature})
is the domain of integration: A disc in Brouwer's formula, and a
cylinder here. However, since $\E (0,t)=0$ the bottom of the
cylinder may be pinched to a point and the cylinder turns to a
disk. The r.h.s. is the trace of the curvature of the connection
$\hat P_jd$ (see e.g. \cite{ref:avron}, Sect. 9.5) or of its
connection 1-form $i[(dS)S^*]_{jj}=\E_{jj} dt-{\cal T}_{jj} dE$.
Similar equations are found in the context of the quantum Hall
effect \cite{ref:stone}, see e.g. \cite{ref:avron} Sect. 11.3), but, unlike there,
$\int^\mu_{E_-} \oint dE\wedge dt\,\Omega(E,t)_{jj}$ is not a
Chern number as a rule, since the integration manifold has a
boundary.

\subsection{The two channel case}
The two channel case is particularly simple. $\ket{\psi}= \begin{pmatrix}
\mathrm{r} \\
 \mathrm{t'}
\end{pmatrix}$ lives in $S^3$. The projection associated to
$\ket{\psi}$ can be identified with a point on $S^2$ according to
\begin{equation}
\ket{\psi}\bra{\psi}= \frac{1+\hat n\cdot\vec\sigma}2
\end{equation}
where $\hat n$ is a unit vector in $\mathbb{R}^3$ and $\vec\sigma$
is the triplet of Pauli matrices. Eq.~(\ref{eq:q-curvature}) then
says that the charge transport is half the spherical angle.

The claim that one can compute the charge transport in a closed
cycle while forfeiting all knowledge of the global phase means, in
the 2-channel case, that all that matters is
$z=\mathrm{r}/\mathrm{t'}$. $z$ lives in ${\bf CP}^1 =
\mathbb{C} \cup \{\infty\}$ and is related to
$\hat n$ above via the usual stereographic projection
 \begin{equation}
\ket{\psi} \to (2 \Re\mathrm{(r \bar t')}, 2 \Im \mathrm{(r \bar t')}, \mathrm{|r|^2
- |t'|^2)} = \left ({2 \Re(z) \over 1 + |z|^2}, {2 \Im(z) \over 1 +
|z|^2}, {|z|^2 -1 \over 1 + |z|^2}\right), \end{equation}
The curvature
\begin{equation}
i ( \mathrm{d\bar{r} dr + d \bar{t'} dt'}) =
i(1+|z|^2)^{-2} d\bar z dz,
\label{eq:z-curvature}
\end{equation}
can be written entirely in terms of
$z$. The current pumped by a small loop in parameter space can
therefore be computed by a calculation on $S^2$.

For large loops, however, things are more subtle. If a path in
parameter space maps to the equator, is the bounding region the
northern hemisphere (area $+ 2\pi$) or the southern hemisphere (signed area
$-2 \pi$)? Because of this ambiguity, the path in $S^2$ only
determines the fractional part of the expected charge transported, not
the integral part. In particular, if the path on the unit sphere is
trivial---the ratio $z=\mathrm{r/t'}$ is constant---then the fractional part is zero
and the charge transported is an integer. The bicycle pump, where $\mathrm{t'}$ is
identically zero, is an example of this quantized transport.

How, then, can one determine also the integral part without a
knowledge of the global phase? Consider a region in parameter
space bounded by our path. The integral part is how many times it
wraps around $S^2$ (i.e., the degree of the map). For the bicycle
pump, this is counting how many points on the unit square
$[0,1]\times[0,1]$ correspond to a particular value of $z$, say
$z=0$. In fact, since we know, be elementary arguments, that the
charge transport in each cycle is $n$, there must be (at least)
$n$ points in the interior of the square where the potential is
{\em reflectionless} at the Fermi energy.

\subsection{Chern and winding numbers}
Part of the motivation for pumps is as standards of charge
transport, whence quantization is an issue. In this section we
want to explain why Chern numbers in quantum pumps must vanish.
Instead, in some cases, winding numbers play a role.

The Hopf fibration gives the unit sphere
$S^{2n-1} \subset \mathbb{C}^n$ the structure of a $U(1)$ principal bundle over
${\bf CP}^{n-1}$ via the map
\begin{eqnarray} \pi: S^{2n-1} & \longrightarrow & {\bf CP}^{n-1}\\
|\psi \rangle & \longrightarrow & | \psi \rangle \langle \psi |,
\end{eqnarray}
where we realize ${\bf CP}^{n-1}$ as the set of all rank-1 projections in
$\mathbb{C}^n$.

Chern numbers in transport theory \cite{ref:thoulessbook} typically arise as
follows. The physical parameter space is a closed surface,
${\cal M}$, say a sphere or a torus. The Hamiltonian, acting on the vector
space ${\bf C}^n$, is then a function on ${\cal M}$, and any one of its
simple eigenvalues
map ${\cal M}$ to ${\bf CP}^{n-1}$. The pullback of the Hopf fibration is
then a $U(1)$ bundle over parameter space, with local geometry and
possibly nontrivial topology. The integrated curvature is a Chern
number, and may be nonzero.

In adiabatic pumps, however, the frozen $S$ matrix defines a map from
${\cal M}$ directly to $S^{2n-1}$, and indirectly to ${\bf CP}^{n-1}$.
This gives a trivialization of the bundle, and shows that all
Chern numbers are zero. Besides, as seen from Eq.~(\ref{eq:q-curvature}),
charge transport is the integral of the Chern character
$i\bra{d\psi_j}d\psi_j\rangle$ over a surface $D$ with boundary, and thus not
given by a Chern number.


More appropriate is the point of view on pumps taken in the
previous section, or its generalization to $n$ leads. If a pump
operation is of the form (\ref{eq:grcirc}), then $\oint\average{
dQ}$ is the winding number of $e^{i\gamma}$ \cite{ref:kamenev,
ref:AEGS:PRL}. Alternatively, the curvature
Eq.~(\ref{eq:p-curvature}) can be computed in terms of $\hat
P_j=\pi(|\psi\rangle)$, cf. Eq.~(\ref{eq:z-curvature}), and the
integral (\ref{eq:q-curvature}) thus performed over $\pi(D)$. The
condition (\ref{eq:grcirc}) means $\partial\pi(D)=\varnothing$,
which again shows that $\oint \average{ dQ}$ is an integer.

An example where this quantization occurs is the U-turn pump,
section \ref{se:Uturn}: The charge transport in a cycle is the
winding number of the map from the circle of fluxes to the
circle of complex numbers of modulus one. A small change of the
parameters underlying Fig.~\ref{qhall} will only modify $k_F\ell$
in the effective description by Eq.~(\ref{eq:hall-scatt}) and thus
preserve quantization. On the other hand, if the U-turn pump is
viewed as an example of a Schr\"odinger operator on a graph, then
a generic perturbation in this class will destroy quantization.



\subsection{Geometry of dissipation and noise}
\label{se:diss&noise}

Eq.~(\ref{eq:dissipation}) and Eq.~(\ref{eq:entropyflow}) that
describe dissipation and entropy (and hence also noise) currents
have a simple geometric interpretation in terms of the fiber bundle
$S^{2n-1} \to {\bf CP}^{n-1}$. The $j$-th row of $\E $ describes the
velocity of $\ket{\psi}$ in $S^{2n-1}$. Of this, ${\E } _{jj}$ is
the projection of this velocity onto the fiber, and ${\E } _{jk}$,
with $k \ne j$, give the projection of this velocity onto ${\bf CP}^{n-1}$.
The current $\average{\dot Q_j}$, and the minimal dissipation
$|{\E } _{jj}|^2/4\pi$, are both functions of motion in the fiber,
while the excess dissipation
\begin{equation}
\average{\dot E_j} - \mu\average{ \dot Q_j} - \pi \average{\dot Q_j}^2
= {1 \over 4\pi}
\sum_{k \ne j} |{\E } _{jk}|^2
\label{excdiss}
\end{equation}
is the ``energy'' (that is, squared velocity)
associated with motion in the base. In particular, a pump operation
is of the form (\ref{eq:grcirc}) if and only if the bound
Eq.~(\ref{eq:optimal}) is saturated, or equivalently if the noise at $T>0$
or at $T=0$, see Eqs.~(\ref{eq:entropyflow}, \ref{eq:shot-noise}), vanishes.
Such pumps may be called {\em optimal} \cite{ref:AEGS:PRL} w.r.t. the $j$-th
channel.

In an interesting piece of work Mirlin and Makhlin
\cite{ref:MakhlinMilrin}, relate the problem of finding a cycle
with minimal noise production at $T=0$ to the problem of finding
a minimal surfaces supported by a given loop. This result is
outside the scope of transport properties which are local in time, as
it deals with $T=0$ and the noise.\\

{\bf Acknowledgment} JEA thanks I. Klich for many enlightening
discussions on the Leso\-vik-Levitov formula. We thank
M. B\"uttiker, J. Fr\"ohlich,
M. Hager, B. Pedrini, K. Schnee, W. Zwerger for discussions.
This work is supported by the
Technion fund for promotion of research and by the EU grant
HPRN-CT-2002-00277.
\appendix

\section{Comparison with the theory of full counting statistics}
\label{sec:levitov}

The formulas for the entropy and noise currents,
Eq.~(\ref{eq:entropyflow}), are singular in the limit $T\to 0$.
This appears to be in conflict with the Lesovik-Levitov's formula (LL) which
gives finite noise at $T=0$ (see below). We shall verify here that, in
fact, for $T\gtrsim\omega$, LL is consistent with our
results. We shall start by recalling LL.

\subsection{The Lesovik-Levitov formula}
Here we shall describe a version of the Lesovik-Levitov formalism
\cite{ref:Klich} where LL is an identity rather than
an (adiabatic) approximation.

Assume that in the distant past and distant future the Hamiltonian
$H(t)$ of the pump coincides with $H_0$, the disconnected pump.
Let $Q_j$ be a projection on one the $j$-th channels. Since the
channel is fixed we suppress the index $j$ below. It is important
in this formulation that the channels are disconnected at the
distant past and distant future and that $[Q,H_0]=0$. It is also
assume that the initial state of the system is thermal state
$\rho(H_0)$. The counting statistics can be described by
means of the characteristic function
$\chi(\lambda)=\sum_{n=-\infty}^\infty p_n e^{i\lambda n}$, where
$p_n$ is the probability for $n$ charges having been transferred to
channel $j$ in the course of whole process. The formula is
\begin{equation}\label{eq:kl}
\chi(\lambda)= \det\bigg(1+\rho \big(e^{-i\lambda
Q/2}\,e^{i\lambda \hat Q}\,e^{-i\lambda Q/2}-1\big)\bigg),\quad
\hat  Q=S^*_dQS_d,\quad \rho=\rho(H_0)
\end{equation}
where, as before, $\rho$ is the Fermi function and $\hat Q$ is a
projection on the states feeding the channel in question.
\begin{rem} The reason $Q$ plays a special role is traced to the
fact that the second quantized $d\Gamma(Q)$ is the number (charge)
operator in the channel. \end{rem} Since $Q$ is a projection
$e^{i\lambda Q}=1+(e^{i\lambda}-1)Q$. Using this one finds
\begin{eqnarray}\label{eq:kla}
e^{-i\lambda Q/2}\,e^{i\lambda \hat  Q}\,e^{-i\lambda Q/2}-1&=&
i\,A\,\sin\lambda -(1-\cos\lambda) A^2\\ &+&
2i\sin(\lambda/2)(1-\cos(\lambda/2)) (Q\hat  QQ_\perp+Q_\perp \hat  Q
Q)\nonumber
\end{eqnarray} where
\begin{equation}\label{eq:A-def}
A=\hat  Q-Q
\end{equation}
is a difference of projections. To second order in $\lambda$
\begin{eqnarray}\label{eq:kl-log-exp}
\log\chi(\lambda)&=& \Tr\log\bigg(1+i\,\lambda\rho\,A
-\frac{\lambda^2}2\rho\,A^2\bigg)+O(\lambda^3)\\
&=& i\lambda\, \Tr\big(\rho\,A\big)- \frac{\lambda^2}2\,
\Tr\big(\rho\,A\big(1-\rho\big)A\big)+O(\lambda^3) \nonumber
\end{eqnarray} In the last step we used the fact that $Q$ commutes
with $H_0$ and the cyclicity of the trace.
\subsection{Charge transport}
The expectation value of charge transport into the $j$-th channel is the
first moment:
\begin{equation}\label{eq:charge}
\average{Q}_j=-i(\log\chi)'(0)=\Tr\big(\rho\,A\big)=
\Tr\big(\big(\rho(H_0-\E_d)-\rho(H_0)\big)Q_j\big)
\end{equation}
This is clearly the correct result, independent of the adiabatic
limit, for the rhs is precisely what one means by the change in
the total charge in the $j$-th reservoir. In the adiabatic limit $\E_d$
is small and hence
\begin{equation}
\average{Q}_j\approx- \Tr\big(\rho'(H_0)\E_dQ_j\big)=-
\frac 1 {2\pi}\int dt\, dE\, \rho'(E)\E_{jj} (E,t)
\end{equation}
in agreement with BPT. In the last step we used
Eq.~(\ref{eq:prod-trace-symbol}) and Eq.~(\ref{eq:rho&symbol}).

\subsection{Splitting the noise }

Noise is the variance of the distribution associated to
$\chi(\lambda)$ or, more precisely, the variance per unit time.
It splits into two positive terms. One term is
proportional to the temperature --- this is the Johnson-Nyquist
noise. The second term involve correlations at different times and
survives at $T=0$. This is the quantum shot noise.

The variance is
\begin{equation}\label{eq:noise}
\average{(\Delta Q)^2} =-  (\log\chi)^{\prime\prime} |_{\lambda  =
0}= \Tr\big(\rho A\big(1-\rho\big)A\big),
\end{equation}
Now, write
\begin{equation}\label{eq:gym}
\rho A(1-\rho)A=\rho A^2(1-\rho)+\rho A[A,\rho] =\rho (1-\rho)
A^2+\rho [\rho,A]A.
\end{equation}
Using the cyclicly of the trace and the average of the two terms
in Eq.~(\ref{eq:gym}), we find
\begin{eqnarray}\label{eq:trace} \Tr\big(\rho
A(1-\rho)A\big)&=& \Tr \big(\rho
(1-\rho)\,A^2\big)+\half\,\Tr\big([\rho,
A]\,[A,\rho]\big).
\end{eqnarray}
Each term is positive.
The Johnson-Nyquist noise is the first term
\begin{equation}\label{eq:nyquistnoise}
Q^2_{JN}=\Tr\big(\rho (1-\rho)\,A^2\big)=-T\,
\Tr(\rho^\prime A^2)\ge 0,
\end{equation}
and the quantum shot noise is the second term:
\begin{equation}\label{eq:jn+qs}
Q^2_{QS}=\half\,\Tr\big([\rho, \hat  Q]\,[\hat
Q,\rho]\big)=\half\,\Tr\big([\delta\rho,  Q]\,[
Q,\delta\rho]\big)\ge 0
\end{equation}
(We have repeatedly used $[Q,H_0]=0$). Since the semi-classical
limit of a commutator is of order $\hbar$, the quantum shot noise
vanishes in the classical limit.
\subsection{Thermal noise}
$\rho^\prime$ is a multiplication operator in $E$ which, at low
temperatures, is concentrated near $\mu$. The symbol of $A_j$ is:
\begin{equation}\label{eq:a-symbol}
A_j\Longleftrightarrow a(E,t)= S^*(E,t)P_jS(E,t)-P_j.
\end{equation}
where $P_j$ is an $n\times n$ matrix that projects on the $j$-th
channel. The Johnson-Nyquist noise at low temperatures can be
written as
\begin{eqnarray} \label{eq:variance2}
2\pi \,Q^2_{JN}   &= &
T\int_\mathbb{R} \,dt\,
  \mathrm{ tr}\big(a^2(\mu,t)\big)=
  T\int_\mathbb{R} \,dt\,
  \tr\big(\hat  P(t)+P- \hat  P(t)P-P\hat  P(t)\big) \nonumber \\
  &=& 2T\sum_{k\neq j}\int_\mathbb{R} \,dt\,|S|^2_{jk}(\mu,t),
\end{eqnarray}
where $\tr$ denotes a trace of $n \times n$ matrices,
and we have used the fact that $P$ and $\hat P$ are one
dimensional projections. We see that the Johnson-Nyquist noise is
proportional to the temperature and the time integral of the
conductance at the Fermi energy. It is finite since for large
times $H(t)$ coincides with $H_0$ and the scattering matrix
reduces to the identity.

\subsection{Shot noise at finite temperatures}

The symbol associated to $[\delta\rho,Q]$ is, to leading order,
the matrix
$$-\rho^\prime(E)\big[\E(E,t),P_j\big]$$
where, as before, $P_j$ is the projection on the $j$-th channel.
At finite temperatures $\rho^\prime(E)$ is a smooth function and
it makes sense to look at the square of the symbol as we must by
Eq.~(\ref{eq:jn+qs}).
Now
\begin{equation}\label{eq:[E,Q]squared}
\tr[\E(E,t),P_j\big][P_j,\E(E,t)\big]=2\big((\E^2)_{jj}-(\E_{jj})^2\big)(E,t)
=2\Delta\E_{jj}^2(E,t).
\end{equation}
Substituting in Eq.~(\ref{eq:trace-symbol}) and using the fact
that $\rho^\prime$ is localized near the Fermi energy, we find
\begin{eqnarray}\label{eq:shot}
Q^2_{QS}&=&\frac 1 {2\pi} \int dE\,dt\,\big(\rho^\prime(E)\big)^2
\Delta\E^2_{jj}(E,t)=\frac \beta {2\pi} \int dt \int_0^1 d\rho
\rho(1-\rho) \Delta\E^2_{jj}(\mu,t)\nonumber \\
&=&\frac{\beta}{12 \pi} \int dt\, \Delta\E^2_{jj}(\mu,t)
\end{eqnarray} in agreement with Eq.~(\ref{eq:entropyflow}).
\subsection{Shot noise at $T=0$}
The result at finite temperature may suggest that the noise over a
pump cycle diverges as $T\to 0$. This is not the case. At $T=0$
the symbol $\rho^\prime$ is a distribution and it is not
permissible to multiply them as we did in Eq.~(\ref{eq:shot}).
However, the problem can be easily avoided by simply not using the
r.h.s. of Eq.~(\ref{eq:jn+qs}) but instead the middle identity. As
before, we approximate the symbol of $\hat Q$ by its value on the
Fermi energy
\begin{equation}\label{eq:q-hat-symbol}
\hat Q_j\Longleftrightarrow S^*(E,t)P_jS(E,t)\approx
S^*(\mu,t)P_jS(\mu,t)=q_j(t)
\end{equation}
This approximation makes $\hat Q_j$ a multiplication operator in
$t$ and hence
\begin{equation}
Q^2_{QS} =  \int_\mathbb{R}\int_\mathbb{R} dt\,
dt'
 \Big|\tilde\rho\left(t-t'\right)\Big|^2 \,
 \tr\big( q(t)- q(t')\big)^2,
\end{equation}
where $\tilde \rho$ is the Fourier transform of the Fermi function.
At $T=0$ the Fermi function associated with chemical potential
$\mu$ is
\begin{equation}\label{eq:rho-tilde}
\tilde\rho(t)=\frac{i}{2\pi (t+i0)}e^{-it\mu}.
\end{equation}
The shot noise at $T=0$, in the limit of large $\mu$, is given by
\begin{equation}
Q^2_{QS}(\mu) = \frac{ 1}{4\pi^2} \int\int dt\, dt'\,\frac{ 1-
\big|\big(S(\mu,t)S^*(\mu,t')\big)_{jj}\big|^2}{(t-t')^2}.
\label{eq:shot-noise}
\end{equation}
Since $S$ is unitary the numerator vanishes quadratically as
$t-t'\to 0$. It follows that the integrand is a bounded function.
However, it is supported on set of infinite area made of two
strips: One along the $t$ axis and one along the $t'$ axis.
Nevertheless, the decay properties of the denominator one easily
sees that the integral is convergent. The noise in a pump cycle
does not diverge as $T\to 0$.



\begin{thebibliography}{10}

\bibitem{ref:akkermans} E. Akkermans, G. Montambaux, Phys. Rev.
Lett. {\bf 68}, 642 (1992); E. Akkermans, J. Math. Phys. {\bf 38},
1781 (1997).


\bibitem{ref:aleiner} L. Aleiner, A.V. Andreev,
Phys. Rev. Lett. {\bf 81}, 1286 (1998).

\bibitem{ref:kamenev} A. Andreev, A. Kamenev,
Phys. Rev. Lett. {\bf 85}, 1294 (2000).

\bibitem{ref:alekseev} A. Alekseev,
{\tt cond-mat/ 0201474}.

\bibitem{ref:avron} J.E.~Avron, {\it Adiabatic Quantum Transport},
Les Houches, E.~Akkermans, et.\ al.\ eds., Elsevier Science (1995).

\bibitem{ref:AEGS:PRB} J.E. Avron, A. Elgart, G.M. Graf, L. Sadun,
Phys. Rev. B {\bf 62}, R10618 (2000).

\bibitem{ref:AEGS:PRL} J.E.~Avron, A.~Elgart, G.M. Graf, L.~Sadun,
Phys. Rev. Lett. {\bf 87}, 236601
(2001).

\bibitem{ref:aegs:JMP} J.E.~Avron, A.~Elgart, G.M. Graf, L.~Sadun,
J.\ Math.\ Phys.\ {\bf 43}, 3415, (2002).

\bibitem{ref:aegss} J. Avron, A. Elgart, G.M. Graf, L. Sadun, K. Schnee,
{\tt math-ph/0209029}.


\bibitem{ref:berry} M.V.~Berry,
Proc. Roy. Soc. London A {\bf 392}, 45 (1984).

\bibitem{ref:Brouwer} P.W. Brouwer,
Phys. Rev. B {\bf 58}, 10135 (1998).

\bibitem{ref:buttiker} M. B\"uttiker,
J. Math. Phys. {\bf 37}, 4793 (1996).

\bibitem{ref:bpt} M. B\"uttiker, A. Pr\^etre, H. Thomas,
Phys. Rev. Lett. {\bf 70}, 4114 (1993); M. B\"uttiker, H. Thomas, A. Pr\^etre,
Z. Phys. B{\bf 94}, 133 (1994).


\bibitem{dolgo} V.T. Dolgopolov, N.B. Zhitenev, A.A. Shashkin,
Pis'ma Zh. Ekp. Theor. Fiz. {\bf 52}, 826 (1990);
JETP Lett. {\bf 52}, 196 (1990).

\bibitem{ref:Wigner} L. Eisenbud, Dissertation, Princeton University, 1948
(unpublished); E.P. Wigner,
Phys. Rev. {\bf 98}, 145 (1955).

\bibitem{ref:friedel} J. Friedel,
Philos. Mag. {\bf 43}, 153 (1952);
L. D. Landau, E. M. Lifshitz, {\it Statistical Mechanics},
Pergamon Press (1978).

\bibitem{fgw}  B.I. Halperin,
Phys. Rev. B {\bf 25}, 2185 (1982);
M. B\"uttiker, 
Phys. Rev. B {\bf 38}, 9375 (1988).


\bibitem{ref:Imry} Y. Imry, {\it Introduction to mesoscopic physics}, Oxford
University Press (1997).

\bibitem{ref:Klich} I. Klich, 
{\tt cond-mat/0209642}.


\bibitem{ref:entin} Y. Levinson, O. Entin-Wohlman, P. Wolfle,
{\tt cond-mat/0010494}.

\bibitem{ref:LLL} L.S. Levitov, H. Lee, B. Lesovik,
J. Math. Phys. {\bf 37}, 4845 (1996); D.A. Ivanov, H.W.
Lee, L.S. Levitov,
Phys. Rev. B {\bf 56}, 6839 (1997); L.S. Levitov, M. Reznikov,
{\tt cond-mat/0111057};
L.S.~Levitov,
{\tt cond-mat/0103617}.

\bibitem{ref:semiclassics} R.G. Littlejohn, W.G. Flynn,
Phys. Rev. A {\bf 44}, 5239 (1991).

\bibitem{ref:MakhlinMilrin} Y. Makhlin, A.D. Mirlin,
Phys.\ Rev.\ Lett.\ {\bf 87},
276803 (2001).

\bibitem{ref:MartinSassoli} P.A. Martin, M. Sassoli de Bianchi,
J. Phys. A {\bf 28}, 2403 (1995).

\bibitem{ref:MokbaletButt1} M. Moskalets, M. B\"uttiker,
Phys. Rev. B {\bf 66}, 035306 (2002).

\bibitem{ref:MoskaletButt2} M. Moskalets, M. B\"uttiker,
Phys. Rev. B {\bf 66}, 205320 (2002).  {\tt cond-mat 0208356}.

\bibitem{ref:PoliankiBrouwer} M.L. Polianski, M.G. Vavylov, P.W. Brouwer,
Phys. Rev. B {\bf 65}, 245314, (2002).

\bibitem{ref:robert} D. Robert, {\it Autour de l'approximation
semi-classique}, Birkh\"auser (1987).

\bibitem{ref:schnee} K. Schnee, Dissertation, ETH-Z\"urich, 2002
(unpublished).

\bibitem{ref:sc} P. Sarma, C. Chamon, 
Phys. Rev. Lett. {\bf 87},
096401 (2001) .

\bibitem{ref:simon} B. Simon, {\it Trace ideals and their applications},
Cambridge University Press (1979).


\bibitem{ref:stone} M.~Stone, {\it The Quantum Hall effect}, World Scientific,
Singapore (1992).


\bibitem{ref:marcus} M. Switkes, C.M. Marcus, K. Campman, A.G. Gossard,
Science, 
 {\bf 283}, 1907 (1999).

\bibitem{ref:thouless83} D. Thouless,
Phys. Rev. B {\bf 27}, 6083 (1983); Q. Niu,
Phys. Rev. Lett. {\bf 64}, 1812 (1990).


\bibitem{ref:thoulessbook}D.~J. Thouless, \emph{Topological quantum numbers in
nonrelativistic physics}, World Scientific, Sin\-ga\-pore, 1998.

\bibitem{ref:krein} D.R. Yafaev, {\it Mathematical Scattering Theory}, AMS (1992).

\end{thebibliography}
\end{document}